\documentclass[
reprint,
runinaddress,
nofootinbib,
nobibnotes,
 amsmath,amssymb,
 aps,
floatfix,
]{revtex4-2}

\usepackage{graphicx}% Include figure files
\usepackage{dcolumn}% Align table columns on decimal point
\usepackage{bm}% bold math
\usepackage{amsmath,amssymb,amsfonts}%
\usepackage{amsthm}%
\usepackage{mathrsfs}%
\usepackage{xcolor}%

\usepackage{dsfont}

\begin{document}

%\preprint{APS/123-QED}

\title{Quantum Control of an Oscillator with a Kerr-cat Qubit}% Force line breaks with \\

\footnotetext[1]{These authors contributed equally to this work.}

\author{Andy Z. Ding$^{1}$}
\email{zhenghao.ding@yale.edu}

\author{Benjamin L. Brock$^{1}$}
\email{benjamin.brock@yale.edu}

\author{Alec Eickbusch} \thanks{Present address: Google Quantum AI, Santa Barbara, CA}
\author{Akshay Koottandavida}
\author{Nicholas E. Frattini} \thanks{Present address: Nord Quantique, Sherbrooke, QC J1J 2E2, Canada}
\author{Rodrigo G. Corti\~{n}as}
\author{Vidul R. Joshi}
\author{Stijn J. de Graaf}
\author{Benjamin J. Chapman} \thanks{Present address: Microsoft Azure Quantum}
\author{Suhas Ganjam} \thanks{Present address: Google Quantum AI, Santa Barbara, CA}
\author{Luigi Frunzio}
\author{Robert J. Schoelkopf}
\author{Michel H. Devoret}\email{michel.devoret@yale.edu}

\affiliation{Departments of Applied Physics and Physics, Yale University, New Haven, CT, USA}

\date{\today}% It is always \today, today,

\begin{abstract}

Bosonic codes offer a hardware-efficient strategy for quantum error correction by redundantly encoding quantum information in the large Hilbert space of a harmonic oscillator. However, experimental realizations of these codes are often limited by ancilla errors propagating to the encoded logical qubit during syndrome measurements. The Kerr-cat qubit has been proposed as an ancilla for these codes due to its theoretically-exponential noise bias, which would enable fault-tolerant error syndrome measurements, but the coupling required to perform these syndrome measurements has not yet been demonstrated. In this work, we experimentally realize driven parametric coupling of a Kerr-cat qubit to a high-quality-factor microwave cavity and demonstrate a gate set enabling universal quantum control of the cavity. We measure the decoherence of the cavity in the presence of the Kerr-cat and discover excess dephasing due to heating of the Kerr-cat to excited states. By engineering frequency-selective dissipation to counteract this heating, we are able to eliminate this dephasing, thereby demonstrating a high on-off ratio of control. Our results pave the way toward using the Kerr-cat to fault-tolerantly measure error syndromes of bosonic codes.

\end{abstract}

\maketitle

A major obstacle to scaling up quantum computers is noise, which causes logical errors and prevents the reliable execution of quantum algorithms. Quantum error correction (QEC) provides a path toward fault-tolerance \cite{Shor1996, Knill1998, Aharonov2008, Kitaev2003}, but this typically comes at the cost of significant resource overhead, often requiring hundreds of physical qubits per logical qubit \cite{Fowler2012, Kivlichan2020, Lee2021, Gottesman2014, Breuckmann2021}. Bosonic codes offer a hardware-efficient alternative to multi-qubit QEC codes by redundantly encoding quantum information in the large Hilbert space of a harmonic oscillator \cite{CLY1997, GKP2001, Michael2016, Mirrahimi2014, Joshi2021,Cai2021}. These codes have been employed to achieve landmark experimental demonstrations, including beyond break-even QEC of quantum memories \cite{Ofek2016,Sivak2023,Ni2023} and fully-autonomous QEC protocols \cite{Gertler2021,deNeeve2022,Lachance-Quirion2024}. However, these experiments rely on an ancilla qubit to control the oscillator and perform quantum error correction, such that errors on the ancilla can propagate to the logical qubit, limiting its lifetime. 

In circuit quantum electrodynamics \cite{Blais2004, Blais2021} these realizations of bosonic codes typically use a microwave cavity as the oscillator and a transmon as the ancilla.  The two are coupled dispersively, and since this dispersive interaction is transparent to transmon phase-flip errors, these QEC protocols are usually only sensitive to transmon bit-flip errors. Although fault-tolerant error syndrome measurements have been experimentally demonstrated using a transmon ancilla \cite{Rosenblum2018}, another approach to achieving fault-tolerance is to use a biased-noise qubit as an ancilla for bosonic codes \cite{Puri2019}. Ideally, the error channel of such an ancilla should be dominated by phase flips, with a negligible rate of bit flips compared to other rates in the system. 

The Kerr-cat qubit (KCQ) has the potential to be exactly such an ancilla due to its promise of an exponential noise bias \cite{Puri2017}. Recent experimental realizations of KCQs have reached a strong noise bias of about $1000$  \cite{frattini2022, Hajr2024} (corresponding to a bit-flip lifetime of $\sim 1$ ms and phase-flip lifetime of $\sim 1$ $\mu$s), and although this is not exponentially large there are many possible methods for further improvement \cite{Putterman2022, Bhandari2024, venkatraman2024, Gravina2023, Ruiz2023}. With a strong noise bias and fast single-qubit gates \cite{Grimm2020,Hajr2024}, the only remaining ingredient for using the KCQ as an ancilla for bosonic codes is an entangling operation between the KCQ and an oscillator that enables the measurement of error syndromes.

In this work, we experimentally demonstrate a coherent parametrically-driven conditional displacement (CD) gate between a KCQ and a high-quality-factor microwave cavity, where the cavity is displaced in one of two directions depending on the state of the KCQ. Combined with single-qubit gates on the KCQ, this CD gate enables universal quantum control of the cavity \cite{Eickbusch2022}. We use this CD gate to measure the decoherence of the cavity in the presence of the KCQ and discover excess cavity dephasing due to heating of the KCQ into excited states, an effect that was not previously predicted. However, by engineering frequency-selective dissipation to counteract this heating \cite{Putterman2022}, we are able to eliminate this dephasing up to the precision of our measurements. This lack of dephasing indicates that the two systems do not entangle unless we are actively driving their interaction, demonstrating a high on-off ratio of control. Our results pave the way toward using the Kerr-cat as an ancilla for fault-tolerant syndrome measurements of bosonic codes \cite{Puri2019}, in particular the Gottesman-Kitaev-Preskill code \cite{GKP2001} whose error syndromes can be mapped to an ancilla via CD gates \cite{Terhal2016, Campagne2020, Fluhmann2020}. 

%In addition to our demonstration of quantum control, our discovery of KCQ-induced cavity dephasing and our implementation of frequency-selective dissipation to mitigate this effect will be useful for understanding how to couple KCQs to other systems as well.

%Our motivation for coupling the KCQ to a cavity, in particular, is to eventually use this setup to fault-tolerantly stabilize a Gottesman-Kitaev-Preskill qubit in the cavity \cite{Puri2019, GKP2001}.  Our discovery of KCQ-induced cavity dephasing and our implementation of frequency-selective dissipation to mitigate this effect will be useful for understanding how to couple KCQs to other systems as well.

\begin{figure}
%\centering
\includegraphics[width=0.47\textwidth]{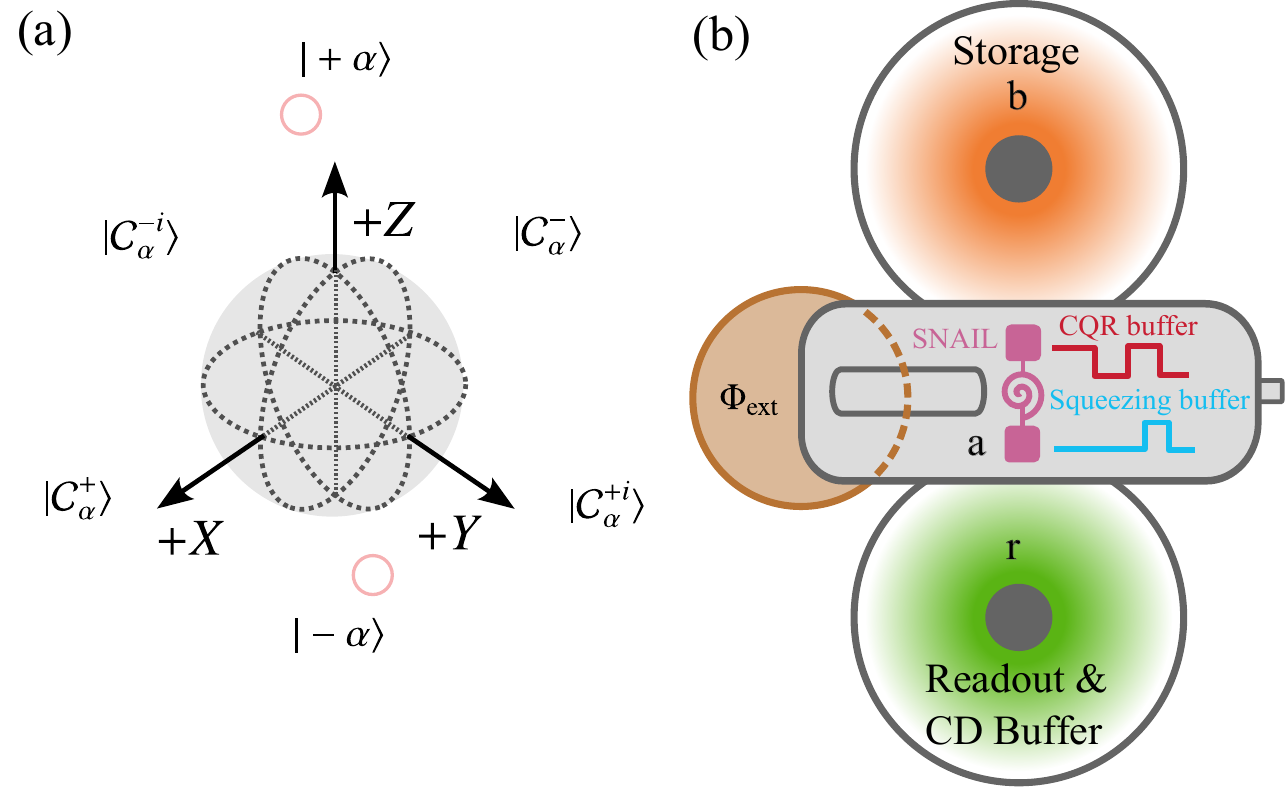}
\caption{(a) Bloch sphere of the Kerr-cat qubit (KCQ). (b) Schematic of the experimental setup for coupling a KCQ (realized in a SNAIL mode, pink) to a high-Q storage cavity. The active modes of the system are the storage mode, the readout mode and the SNAIL (Superconducting Nonlinear Asymmetric Inductive eLement) \cite{frattini2017, Sivak2019, FrattiniThesis} mode, whereas the buffer modes are used to facilitate parametric drives to the SNAIL, in particular the CQR (cat qubit readout) drive, the CD (conditional displacement) drive and the squeezing drive.}\label{fig1}
\end{figure}

The KCQ consists of the degenerate ground state manifold of the Hamiltonian 
\begin{equation}
    H_{\mathrm{KCQ}}/\hbar = -Ka^{\dagger 2}a^2 + \epsilon_2 a^{\dagger 2} + \epsilon_2^* a^2,
    \label{Eq.KCQ}
\end{equation}
where $a$ is the annihilation operator of the Kerr-oscillator mode, $K$ is the Kerr nonlinearity of the mode, and $\epsilon_2$ is the strength of the squeezing drive. This computational subspace is spanned by the orthornormal even- and odd-parity cat states $| \mathcal{C}^\pm_\alpha \rangle = \mathcal{N}_\alpha^\pm (|+\alpha \rangle \pm |-\alpha \rangle)$ (with normalization $\mathcal{N}_\alpha^\pm = 1/\sqrt{2\left( 1 \pm e^{-2\alpha^2}\right)}$) \cite{Puri2017}, where $\alpha = \sqrt{\epsilon_{2}/K}$ is assumed to be real without loss of generality. This gives rise to the Bloch sphere depicted in Fig. \ref{fig1} (a). These cat states $| \mathcal{C}^\pm_\alpha \rangle$ form the X basis of the KCQ, while the parity-less cats $| \mathcal{C}^{\pm i}_\alpha \rangle = \mathcal{N}_\alpha^\pm (|+\alpha \rangle \pm i|-\alpha \rangle)$ form the Y basis. The Z basis consists of the even and odd superpositions of $| \mathcal{C}^\pm_\alpha \rangle$, which are approximately equal to the coherent states $|\pm \alpha \rangle$ for large $\alpha$. The noise bias of the KCQ comes from the metapotential associated with $H_\mathrm{KCQ}$ \cite{dykman_fluctuating_2012, Puri2019}: it has a double-well structure with two global minima (corresponding to the coherent states) separated by a potential barrier \cite{Grimm2020}. Dephasing of the Kerr-oscillator causes tunneling events between the wells, corresponding to bit-flip errors, which are exponentially suppressed by a factor $2\alpha^{2}\exp(-2\alpha^{2})$ due to the height of the potential barrier \cite{venkatraman2024}. On the other hand, photon-loss events in the Kerr-oscillator cause the cat states to change parity, corresponding to phase-flip errors, which are linearly amplified by a factor $2\alpha^{2}$ due to the average number of photons in the coherent states \cite{Mirrahimi2014, Puri2017, Grimm2020, frattini2022}.

Following previous demonstrations \cite{Grimm2020, frattini2022, venkatraman2024, Hajr2024}, we experimentally realize the KCQ by squeezing a capacitively-shunted Superconducting Nonlinear Asymmetric Inductive eLement (SNAIL) \cite{frattini2017, Sivak2019, FrattiniThesis}. The SNAIL is a flux-tunable circuit described by the Hamiltonian $H_{\mathrm{SNAIL}}/\hbar = \omega_a (\Phi_\mathrm{ext}) a^{\dagger} a + g_3 (\Phi_\mathrm{ext}) (a + a^\dagger)^3 + g_4 (\Phi_\mathrm{ext}) (a + a^\dagger)^4 $, where $g_3$ and $g_4$ are the third and fourth order nonlinearities of the SNAIL, and $\Phi_\mathrm{ext}$ is the external flux threading the SNAIL. The KCQ Hamiltonian in Eq. \ref{Eq.KCQ} is obtained as the static effective Hamiltonian of the system when we apply a squeezing drive to the SNAIL at frequency $\omega_s = 2\omega_a$. A continuous $\sigma_z (\theta)$ rotation of the KCQ is realized by driving the SNAIL at its resonant frequency $\omega_s/2 = \omega_a$, while a discrete $\sigma_{x}(\pi/2)$ rotation is realized by turning off the squeezing drive and allowing the SNAIL mode to evolve under its bare Kerr-nonlinear Hamiltonian for time $T_K = \pi/K$ \cite{Kirchmair2013, Grimm2020}. Together, these rotations enable universal single-qubit gates on the KCQ. To read out the logical information stored in this qubit, we use a process called cat-qubit readout (CQR) \cite{Grimm2020}, which involves parametrically driving a beam-splitter interaction between the SNAIL circuit and a readout cavity (frequency $\omega_r$ and mode operator $r$) at the difference frequency $\omega_r - \omega_s/2$ while stabilizing the KCQ. After projecting the Hamiltonian of this interaction onto the computational basis of the KCQ, it takes the form of a conditional displacement of the readout cavity $H_{\mathrm{CQR}} = g_{\mathrm{CQR}}(r+r^{\dag})\sigma_z$ \cite{Grimm2020}, such that measuring the phase of the radiation leaking out of the readout cavity constitutes a readout of the KCQ along its Z axis.

The same conditional displacement interaction can be activated between the KCQ and a high-Q storage cavity (frequency $\omega_b$ and mode operator $b$) by driving the SNAIL at the difference frequency $\omega_b - \omega_s/2$. For clarity, we reserve the acronym CD for the conditional displacement interaction between the KCQ and the high-Q cavity. While driving this beamsplitter interaction at a rate $g_\mathrm{BS}$, the static effective Hamiltonian of the system in the rotating frame of the SNAIL and the storage cavity takes the form \cite{Grimm2020}
\begin{equation}
    H/\hbar = H_\mathrm{KCQ}/\hbar - \chi_{ab} a^{\dagger}a b^{\dagger}b + g_{\mathrm{BS}}a^\dagger b + g^*_{\mathrm{BS}} ab^\dagger,
\end{equation}
where $\chi_{ab} a^{\dagger}a b^{\dagger}b$ is the parasitic cross-Kerr interaction and $g_{\mathrm{BS}}a^\dagger b + g^*_{\mathrm{BS}} ab^\dagger$ is the beamsplitter interaction. Assuming that the KCQ remains in its degenerate ground state manifold, we project onto the cat-qubit subspace with the projector $\mathcal{P}_\mathcal{C} = | \mathcal{C}^+_\alpha \rangle \langle \mathcal{C}^+_\alpha | + | \mathcal{C}^-_\alpha \rangle \langle \mathcal{C}^-_\alpha |$ and obtain \cite{sp}
\begin{equation}
\begin{split}
    H_{\mathrm{int}}/\hbar = & g_{\mathrm{BS}} \alpha \left( \left( b^\dagger + b \right) \sigma_z + ie^{-2\alpha^2} \left( b^\dagger - b \right) \sigma_y \right) \\
    & -\chi_{ab} \alpha^2 b^\dagger b \left( \mathds{1} - 2e^{-2\alpha^2} \sigma_x \right).
\end{split}
\label{Eq.2}
\end{equation}
In general, the cross-Kerr interaction will lead to dephasing of the storage cavity due to photon shot noise in the SNAIL. As we increase $\alpha$, however, the entangling term proportional to $b^{\dagger}b\sigma_{x}$ becomes exponentially suppressed such that the cross-Kerr interaction simplifies into a Stark shift, thereby preventing KCQ-cavity entanglement during idling time. This feature is crucial for engineering an ancilla with a large on-off ratio of cavity control, where the interaction between the ancilla and the cavity is only activated via a driven parametric process. 

In the limit of $\alpha \gg 1$, Eq. \ref{Eq.2} simplifies to a conditional displacement interaction and a Stark shift given by
\begin{equation}
    H_{\mathrm{int}}/\hbar \approx g_{\mathrm{BS}} \alpha \left( b^\dagger + b \right) \sigma_z -\chi_{ab} \alpha^2 b^\dagger b.
    \label{Eq.CD}
\end{equation}
By tracking this Stark shift and evolving for time $t = \beta /(2g_\mathrm{BS}\alpha)$, we realize the conditional displacement unitary $\mathrm{CD}(\beta) = \mathrm{D}(-\beta/2)|-\alpha\rangle \langle -\alpha| + \mathrm{D}(+\beta/2)|+\alpha\rangle \langle +\alpha|$, where $\mathrm{D}(\beta)=\exp(\beta b^\dagger - \beta^* b)$ is the displacement operator and the conditional displacement rate $g_\mathrm{CD} = 2g_{\mathrm{BS}} \alpha$ is enhanced by having more photons in the KCQ. Together with the single qubit gates on the KCQ, this conditional displacement gate enables universal control of the system \cite{Eickbusch2022}.

A schematic of our experimental setup is shown in Fig. \ref{fig1}(b).  It consists of a sapphire chip sandwiched between two 3D superconducting microwave post cavities machined out of high-purity aluminum \cite{Reagor2013}. On the sapphire substrate we fabricated tantalum-based superconducting circuits \cite{Place2021, Ganjam2024}, which capacitively couple to the 3D cavities via waveguide tunnels. The active modes of our system are the on-chip SNAIL mode ($\omega_a = 2\pi\times 4.00 $ GHz), the fundamental mode of the 3D storage cavity ($\omega_b = 2\pi\times 7.02$ GHz), and the first harmonic of the 3D readout cavity ($\omega_r = 2\pi\times 9.36$ GHz). The remaining modes in our system are buffers for delivering parametric drives to the SNAIL; they are essentially Purcell filters, designed to be detuned from the desired drive frequency by about $50$ MHz \cite{Chapman2023}.  These buffer modes are the on-chip stripline resonators for delivering the CQR drive ($\omega_{\mathrm{CQR}} = \omega_r - \omega_s/2 = 2\pi\times 5.36$ GHz), the on-chip stripline resonator for delivering the squeezing drive ($\omega_{\mathrm{sq}} = 2\omega_a = 2\pi\times 8.00$ GHz), and the fundamental frequency of the 3D readout cavity used for delivering the CD drive ($\omega_{\mathrm{CD}} = \omega_b - \omega_s/2 = 2\pi\times 3.02$ GHz). Magnetic flux is delivered to the SNAIL with a solenoid (consisting of a copper coil wound with superconducting NbTi wire) and an on-chip superconducting flux transformer \cite{Chapman2023}. For all of our experiments we operate the SNAIL at a flux-bias point of $\Phi_\mathrm{ext} = 0.32\Phi_{0}$, where the SNAIL circuit has appreciable $g_3$ and $g_4$ values, and the frequency conditions of the various parametric processes are reasonably close to their buffer modes. See the supplementary material \cite{sp} for more detail on the experimental apparatus and the fabrication process.

With this experimental setup, we have reproduced previously demonstrated KCQ gates and readout, as well as characterized both the bare SNAIL lifetimes and the KCQ lifetimes. In particular, the SNAIL has an energy relaxation lifetime $T_{1,a} = 16.0 \pm 0.4 $ $\mu $s and a Ramsey coherence lifetime $T_{2,a} = 7.2 \pm 0.1 $ $\mu $s. For our standard cat size of $\alpha^2=4$ that is used throughout this manuscript, the KCQ has a coherent state lifetime $T_Z = 147 \pm 4 $ $\mu $s and a cat state lifetime $T_{X,Y} = 2.32 \pm 0.04 $ $\mu $s. The $\sigma_x (\pi/2)$ rotation, realized by free evolution of the SNAIL under its Kerr nonlinearity, takes $T_{K} = 272$ ns corresponding to $K/2\pi = 0.93 \pm 0.03$ MHz. Additional information on our system characterization can be found in \cite{sp}.

\begin{figure}
%\centering
\includegraphics[width=0.4\textwidth]{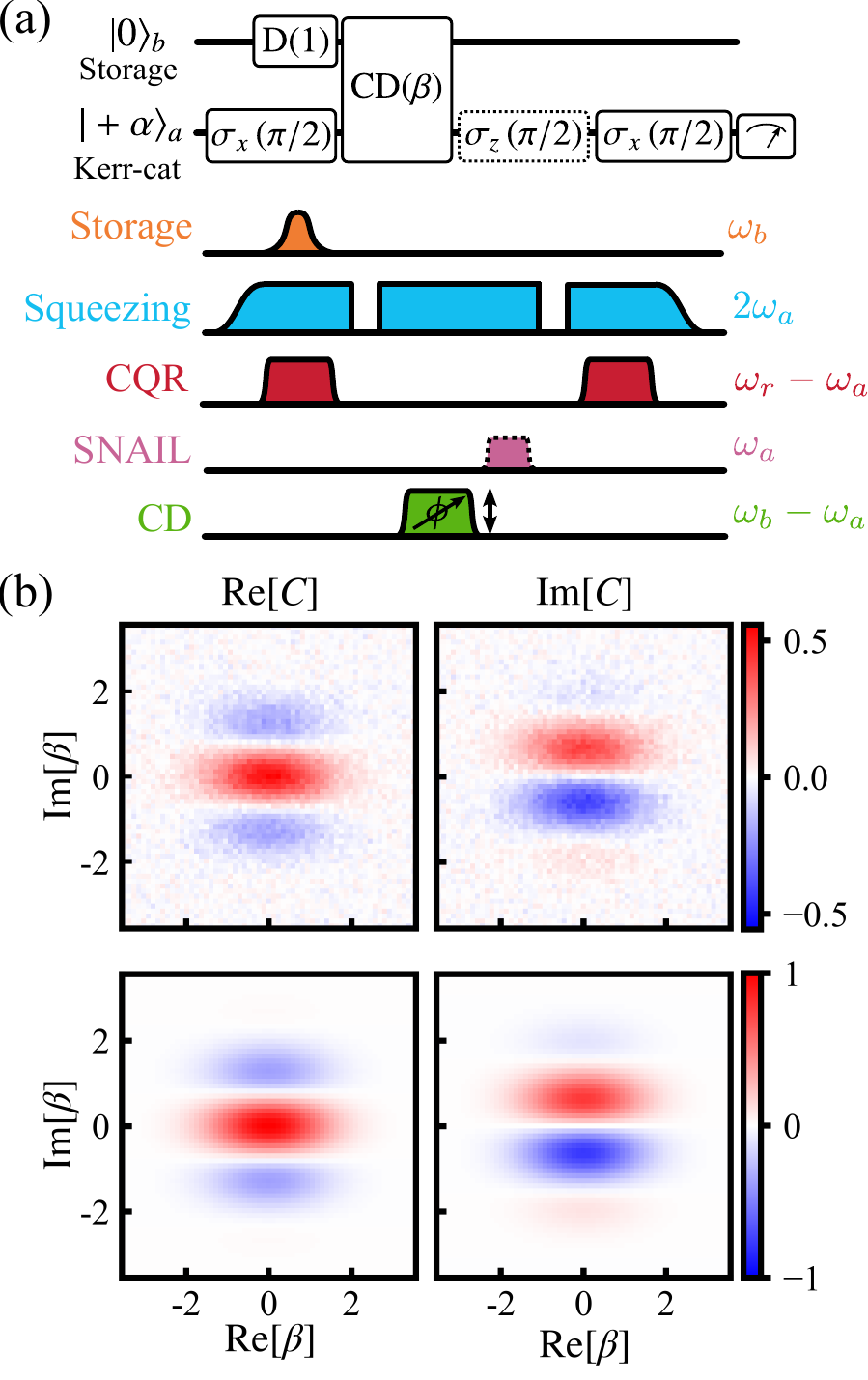}
\caption{(a)  Gate sequence (upper) and pulse sequence (lower) for performing characteristic function tomography of a coherent state of amplitude 1 in the storage cavity with the KCQ. The pulse lengths are not to scale. (b) Experimental (upper) and theoretical (lower) characteristic function tomography of a coherent state with amplitude 1 in the storage cavity, demonstrating the coherence of the conditional displacement interaction. Combined with single-qubit KCQ gates, the CD gate enables universal quantum control of the cavity \cite{Eickbusch2022}. }\label{fig2}
\end{figure}

To demonstrate a coherent conditional displacement interaction between the storage cavity and the KCQ, we perform characteristic function (CF) tomography on the storage cavity using the KCQ \cite{Campagne2020, Fluhmann2020, Eickbusch2022}. The characteristic function is defined as $\mathcal{C} (\beta)=\langle \mathrm{D}(\beta) \rangle$, and it can be measured by initializing the KCQ in the $| \mathcal{C}^{\pm i}_\alpha \rangle$ state, performing $\mathrm{CD}(\beta)$, and measuring the KCQ along its $X$ and $Y$ axes.  The principle of this measurement is that when implementing a conditional displacement $\mathrm{CD}(\beta)$ between the KCQ and the storage cavity, the KCQ rotates about its z-axis by an amount that encodes the phase of the displacement operator, such that $\langle \mathrm{D}(\beta) \rangle = \pm\langle \sigma_y \rangle \mp i \langle \sigma_x \rangle$, where the signs are determined by the choice of initial state.

To perform CF tomography, we follow the pulse sequence depicted in Fig. \ref{fig2}(a). First, we prepare a coherent state of amplitude 1 in the storage cavity by driving the cavity resonantly with a Gaussian pulse (duration $480$ ns).  At the same time, we prepare a parity-less cat state $|\mathcal{C}^{\pm i}_\alpha \rangle$ in the KCQ by ramping up the squeezing drive (duration $2$ $\mathrm{\mu s}$), projecting into a coherent state with a CQR measurement (duration $2.5$ $ \mathrm{\mu s}$), and performing a $\sigma_x (\pi/2)$ gate (duration $272$ ns). Next, we implement a conditional displacement gate $\mathrm{CD}(\beta)$ with varying amplitude and phase of $\beta$ (duration $348$ ns, corresponding to the rate $g_\mathrm{CD}/2\pi=6.18 \pm 0.06 $ MHz). Finally, we repeat the experiment multiple times, measuring the KCQ along the X and Y axes to determine the real and imaginary parts of the characteristic function; to measure along the Y axis we perform a $\sigma_x (\pi/2)$ gate followed by a CQR measurement (duration $ 2.5 $ $\mu$s), whereas to measure along the X axis we prepend this sequence with a $\sigma_z (\pi/2)$ gate (duration $80$ ns).  The results of this measurement are shown in Fig. \ref{fig2}(b), where the contrast of the CF tomography is limited by photon-loss errors in the SNAIL during the $\sigma_x (\pi/2)$ rotation. Aside from reduced contrast, we find excellent agreement between our experimental results and theoretical predictions. Furthermore, the fact that we are able to imprint the phase associated with the displacement operator onto the equator of the KCQ demonstrates the coherence of the CD gate, which we are able to observe due to the CD rate $g_\mathrm{CD}$ being faster than all decoherence rates in our system.  Combined with single-qubit gates on the KCQ, this CD gate enables universal quantum control of the cavity \cite{Eickbusch2022}.

\begin{figure}%[h]%
\includegraphics[width=0.48\textwidth]{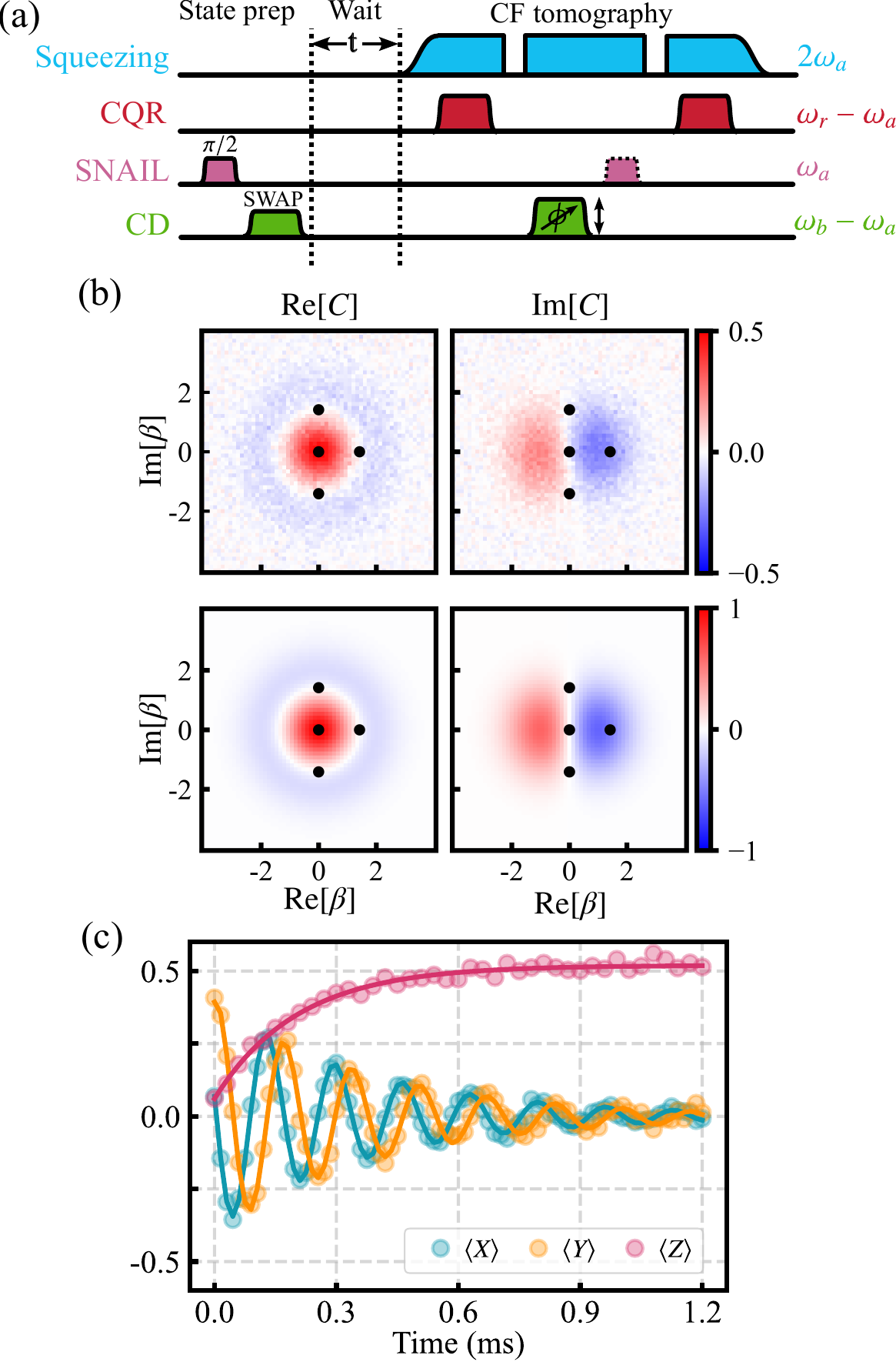}
\caption{(a) Pulse sequence for storage lifetime measurements. We prepare a $(|0\rangle + i|1\rangle)/\sqrt{2}$ state in the storage, wait a variable time $t$, and perform characteristic function tomography at specific values of $\beta$ to reconstruct the expectation values of the Pauli operators of the storage Fock qubit (spanned by the $|0\rangle$ and $|1\rangle$ states). (b) Measured (upper) and predicted (lower) characteristic function (CF) of the $(|0\rangle + i|1\rangle)/\sqrt{2}$ state prepared in the storage cavity. The four points in the plot are the values of $\beta$ at which we sample the CF in order to measure the Pauli expectation values of the Fock qubit in the storage cavity. (c) Evolution of the expectation values $\langle X \rangle$, $\langle Y \rangle$ and $\langle Z \rangle$ of the storage fock qubit as a function of time, from which we can determine $T_1 = 204 \pm 9$ $\mu$s and $T_2 = 381 \pm 8$ $\mu$s. 
}\label{fig3}
\end{figure}

In order to use this platform for bosonic quantum error correction, we need to demonstrate the ability to maintain storage coherence while establishing a KCQ in the SNAIL. From Eq. \ref{Eq.2} and Eq. \ref{Eq.3}, we expect the storage dephasing due to SNAIL photon shot noise to be suppressed as we increase $\alpha^2$, which we can verify experimentally by measuring the energy relaxation rate and dephasing rate of the storage cavity in the presence of a KCQ with varying size $\alpha$. As a first step, we develop a method of measuring the storage coherences that is amenable to our experimental setup.

Similar to other approaches \cite{Fluhmann2020, Campagne2020, Eickbusch2022, Sivak2023, Koottandavida2023}, we measure $T_1$ and $T_2$ of the storage cavity by restricting the cavity to its Fock-qubit subspace (spanned by $|0\rangle$ and $|1\rangle$, with Pauli operators $X$, $Y$, and $Z$), preparing the state $|+Y\rangle = (|0\rangle + i|1\rangle)/\sqrt{2}$, and measuring $\langle X(t) \rangle$, $\langle Y(t) \rangle$ and $\langle Z(t) \rangle$ while the cavity state decays.  Our control sequence for this measurement is shown in Fig. \ref{fig3}(a). To prepare the $|+Y\rangle$ state in the storage we first excite the SNAIL to the $(|g\rangle + i|e\rangle)/\sqrt{2}$ state with a square pulse of duration $1.1 $ $\mu$s, after which we drive a beamsplitter interaction at $\omega_b - \omega_a$ for $1.5 $ $\mu$s to swap the SNAIL state into the storage cavity.  The square pulse of duration $1.1 $ $\mu$s is designed such that its frequency distribution is a sinc function with a notch at the anharmonicity of the SNAIL, thereby minimizing leakage of the SNAIL to its $|f\rangle$ state. To measure $\langle X(t) \rangle$, $\langle Y(t) \rangle$ and $\langle Z(t) \rangle$ we use a method adapted from \cite{Koottandavida2023}: assuming we remain in the Fock-qubit subspace of the storage cavity, these Pauli expectation values can be determined by measuring the characteristic function $\mathcal{C}(\beta)$ at the four points $\beta = {0, \sqrt{2}, \sqrt{2}i, -\sqrt{2}i}$ \cite{sp}.

\begin{figure*}
\includegraphics[width=0.99\textwidth]{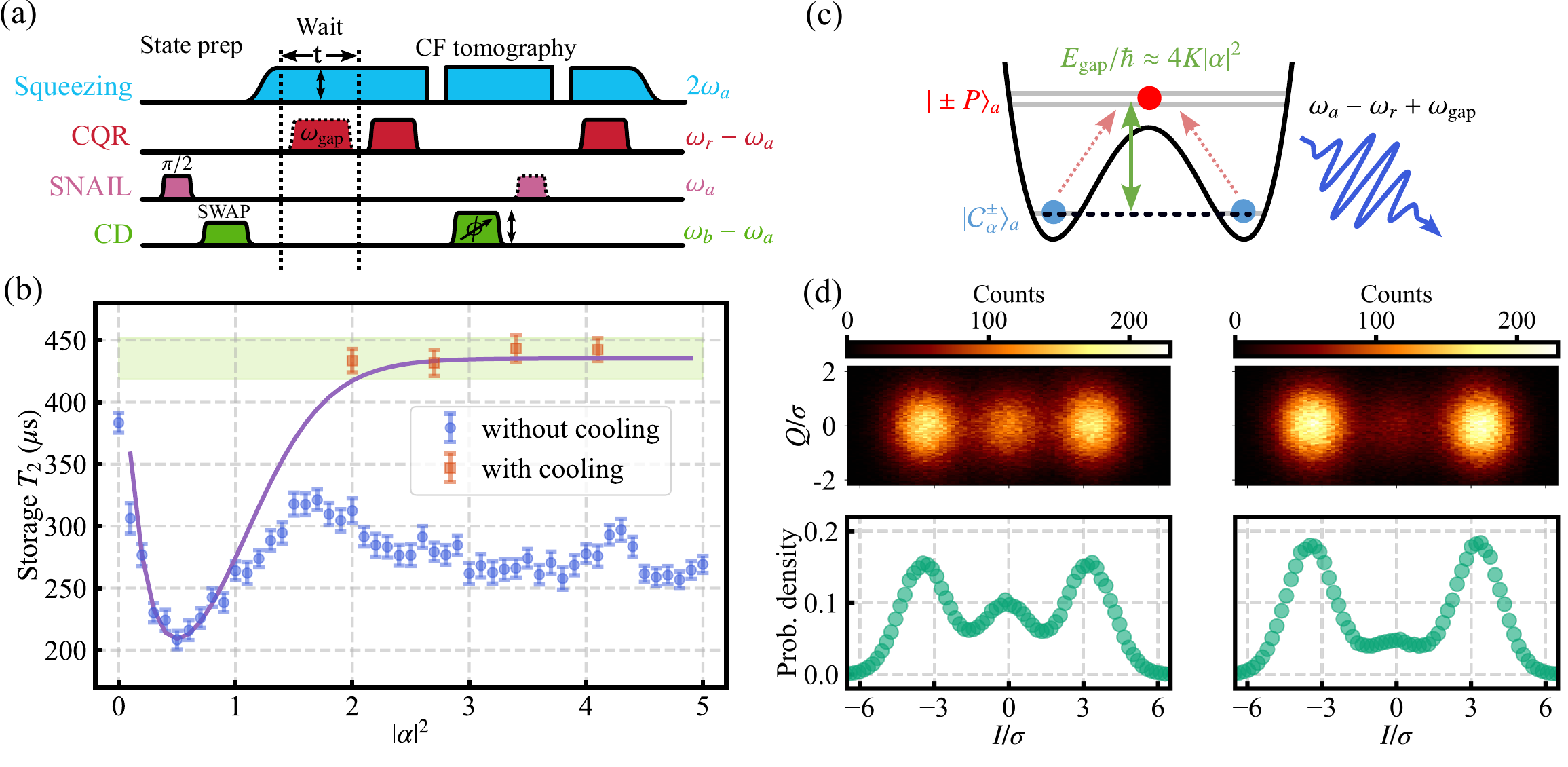}
\caption{(a) Pulse sequence measuring storage $T_1$ and $T_2$ in the presence of a KCQ. (b) Measurement results for storage cavity dephasing in the presence of a KCQ with varying size $\alpha$. The blue circles are measured without cooling the KCQ, and the orange squares are measured with active cooling via frequency-selective dissipation (FSD). The purple line is from a master equation simulation that serves as a guide to the eye for our results \cite{sp}. The green region is our confidence interval of the storage $2T_1$ we obtain from these measurements, taken over a one month period. (c) Schematic representation of the metapotential and the lowest-lying energy levels of a KCQ with $n_{\mathrm{cat}}=4$. We dissipate the excited state population of the KCQ into the readout cavity via a three-wave mixing process. (d) CQR I/Q signal histograms (upper panel) and their marginals along the I quadrature (lower panel) of the KCQ with artificially-induced heating \cite{sp} before (left) and after (right) FSD.}\label{fig4}
\end{figure*}

The experimentally measured $\langle X(t) \rangle$, $\langle Y(t) \rangle$ and $\langle Z(t) \rangle$ are shown in Fig. \ref{fig3}(c), where the decay of $\langle Z(t) \rangle$ tells us $T_1 = 204 \pm 9$ $\mu$s and the ring-downs of $\langle X(t) \rangle$ and $\langle Y(t) \rangle$ tell us $T_2 = 381 \pm 8$ $\mu$s, corresponding to a pure dephasing rate $\Gamma_\phi = (5.7 \pm 2.8 $ $ \mathrm{ms})^{-1}$. Bare microwave cavities have minimal intrinsic dephasing \cite{Reagor2016, Ganjam2024, Milul2023, Chakram2021, Rosenblum2018}, so we expect this dephasing to be dominated by heating in the SNAIL coupling to the storage cavity via the cross-Kerr interaction, which we verify experimentally by measuring the thermal population of the SNAIL. Owing to the low contrast of SNAIL dispersive readout, we cannot carry out this measurement directly \cite{sp}. To get around this limitation, we map the thermal population of the SNAIL into the storage by swapping their states (via a beam-splitter interaction \cite{sp}) after the system reaches thermal equilibrium, and then perform CF tomography on the storage. The width $\sigma_\mathrm{th}$ of the resulting Gaussian is related to the thermal population in the storage according to $\sigma_\mathrm{th}=1/\sqrt{2n_{\mathrm{th}}+1}$. We find $n_{\mathrm{th}}=2.8 \pm 0.5 \%$, corresponding to an induced dephasing rate \cite{Clerk2007} of $\Gamma_\phi = (8.0 \pm 1.6 $ $ \mathrm{ms})^{-1}$, which agrees with our direct measurement of the storage dephasing rate. It is important to note that this method assumes that the storage cavity at $7$ $\mathrm{GHz}$ has a much lower thermal population than the SNAIL at $4$ $\mathrm{GHz}$, but the fact that these two independent measurements of the storage dephasing rate agree with one another is consistent with this assumption.

With this technique for characterizing the storage lifetimes, we can investigate the effect of the second term of Eq. \ref{Eq.2},
\begin{equation}
    H_{\chi}/\hbar = -\chi_{ab} \alpha^2 b^\dagger b \left( \mathds{1} - 2e^{-2\alpha^2} \sigma_x \right),
    \label{Eq.3}
\end{equation}
which arises from the cross-Kerr interaction between the SNAIL and the storage cavity with $\chi_{ab}/2\pi = 2.91 \pm 0.03$ kHz \cite{sp}. The first term is a Stark shift of the storage due to photons in the KCQ, which can be tracked in software. The second term corresponds to a residual cross-Kerr interaction between the KCQ and the storage, where the strength of the interaction peaks at $\alpha^2=1/2$ and is then exponentially suppressed with the number of photons $\alpha^2$ in the KCQ. Under this interaction, parity flips in the KCQ (due to photon shot noise in the SNAIL) induce dephasing in the storage, just as SNAIL heating gave rise to storage dephasing in the previous measurement. When we increase the size of the KCQ, this dephasing is exponentially suppressed, keeping the KCQ disentangled from the storage and preserving the coherence of the storage cavity. To examine this effect experimentally, we modify our previous measurement by stabilizing a KCQ with varying size $\alpha$ while we idle the storage cavity (pulse sequence depicted in Fig. \ref{fig4}(a)). Then we measure $\langle X(t) \rangle$, $\langle Y(t) \rangle$ and $\langle Z(t) \rangle$ of the storage cavity using the method discussed above and shown in Fig. \ref{fig3}. The result of this measurement is shown in Fig. \ref{fig4}(b). As $\alpha$ increases from zero, the cavity experiences more dephasing, causing $T_2$ to decrease. Upon reaching half a photon in the KCQ, we observe maximum storage dephasing and a subsequent revival of storage $T_2$ as $\alpha$ increases, indicating suppression of photon shot noise as predicted in Eq. \ref{Eq.3}. However, the revival does not continue all the way to $T_2=2T_1$, which corresponds to the region colored in light green in Fig. \ref{fig4}(b). This additional channel of dephasing in the storage cavity is due to heating in the KCQ, as we will demonstrate. 

To illustrate the physics of the storage dephasing, we consider a KCQ with $\alpha=2$, the potential energy and eigenstates of which are shown in Fig. \ref{fig4}(c). The states at the minima are the degenerate ground states of the metapotential, forming the computational subspace of the KCQ, whereas the first excited states are non-degenerate excited states outside of the wells. These excited states have different average photon number than the ground states, and therefore cause different Stark-shifts of the storage cavity \cite{sp}. Without heating, the KCQ would populate in the degenerate ground states and remain disentangled from the storage cavity. With heating, indicated by the dotted red arrows, the KCQ would transition from the ground states to the excited states, dephasing the storage cavity.

One way to mitigate this dephasing is through a frequency-selective dissipation (FSD) of the excited state population of the KCQ \cite{Putterman2022}. The idea is to dissipate excited state population through a lossy mode without affecting the ground state manifold. Whereas Ref. \cite{Putterman2022} proposed using filter modes that are on resonance with the energy gap of the KCQ, here we use the three-wave mixing capability of our SNAIL to hybridize the first excited state manifold of the KCQ with the readout cavity. This method is convenient to perform in our setup, as the beamsplitter interaction required by this process is our built-in CQR interaction with a detuning, giving us $\omega_{\mathrm{FSD}} = \omega_r - \omega_s/2 + \omega_{\mathrm{gap}}$. We calibrate the energy gap  $\omega_{\mathrm{gap}}/2\pi = 12.5$ $ \mathrm{MHz}$ at $\alpha=2$ by performing spectroscopy on the excited states of the KCQ \cite{sp}. To calibrate the amplitude of the FSD drive, we need a way to measure the population of the excited state manifold. Due to the fact that the CQR drive does not displace the readout cavity when the KCQ is in an excited state outside of the wells, the center blob in the I/Q plane of the CQR signal correlates with the total excited state population. We set the FSD amplitude to the value that minimizes this signal, as is shown in Fig. \ref{fig4}(d).

With the FSD, we can experimentally verify that the storage dephasing is due to KCQ heating. To do so, we repeat the previous measurement shown in Fig. \ref{fig3} at $\alpha^2=2.0, 2.7, 3.4$ and $4.1$, except now we apply the FSD drive to the KCQ during the storage idling time. As shown in Fig. \ref{fig4}(b), the active cooling on the KCQ improves $T_2$ of the storage cavity to a level consistent with zero dephasing, up to the precision of our measurement, while $T_1$ of the storage cavity is unaffected. In addition to verifying the origin of the storage dephasing, this result validates the interaction Hamiltonian (Eq. \ref{Eq.2}) between the storage and the KCQ: for sufficiently large cats, the storage and the KCQ do not entangle, provided that the KCQ stays in its computational manifold. Furthermore, the fact that the two systems do not entangle unless we are actively driving their interaction demonstrates a high on-off ratio of control.

In summary, we have experimentally realized a coherent parametrically-driven CD gate between a KCQ and a high-quality-factor storage cavity that, combined with single-qubit KCQ gates, enables universal quantum control of the cavity \cite{Eickbusch2022}. By actively cooling the KCQ we were able to prevent the two systems from entangling during idling time, thereby demonstrating a high on-off ratio of control. The operations we have demonstrated constitute all the necessary ingredients for stabilizing Gottesman-Kitaev-Preskill codewords in the storage cavity \cite{GKP2001, Terhal2016, Campagne2020}, paving the way towards fault-tolerant quantum error correction of this code \cite{Puri2019}. To achieve this, we need to implement a higher quality $\sigma_x (\pi/2)$ Kerr-cat rotation, for instance by improving the coherence of the SNAIL or by using alternative strategies for realizing this rotation \cite{Puri2017, Hajr2024, Goto2016}. In addition to performing fault-tolerant syndrome measurements of bosonic codes, the experimental platform we developed for coupling a Kerr-cat to a cavity could be useful for realizing error-transparent gates in hybrid discrete/continuous variable systems \cite{Pietikainen2022}. Finally, our understanding of the Kerr-cat-induced cavity dephasing and its mitigation with active cooling will be important for other experiments in which Kerr-cats are coherently coupled to other systems \cite{Iyama2024,Hoshi2024}. \\

\noindent \textbf{Author contributions} AZD, BLB, and AE designed the experimental setup. AZD fabricated the device, with assistance from SG and LF. SJdG and BJC contributed to the 3D cavity design. AZD and BLB built the microwave control electronics. VRJ designed and fabricated the SNAIL parametric amplifier. AZD performed the measurements with supervision from BLB. AE, AK, NEF, and RGC provided valuable help with the measurements.  BLB and AZD developed the method for performing frequency-selective dissipation on the Kerr-cat. BLB and AK assisted with data analysis and cryogenic maintenance. RJS supervised the contributions of SG, SJdG, and BJC. MHD supervised the project. AZD, BLB, and MHD wrote the manuscript with feedback from all authors.

\noindent \textbf{Ethics declarations} LF and RJS are founders and shareholders of Quantum Circuits Inc. (QCI). The remaining authors declare no competing interests. 

\noindent \textbf{{Data availability}} The data that support the findings of this study are available from the corresponding authors upon a reasonable request. 

\noindent \textbf{Code availability} All computer code used in this study are available from the corresponding authors upon a reasonable request.

\begin{acknowledgments}
We thank S.~M.~Girvin, Y.~Lu, S.~Puri, A.~P.~Read, M.~Sch\"afer, S.~Singh, V.~V.~Sivak, Q.~Su, J.~Venkatraman, and X.~Xu for helpful discussions. We are grateful to J. Curtis for technical assistance and I. Tsioutsios for device fabrication assistance. Finally, we thank Y.~Sun, K.~Woods, L.~McCabe, and M.~Rooks for their assistance and guidance in the device fabrication processes. This research was supported by the US Army Research Office (ARO) under grant W911-NF-23-1-0051 and by the US Department of Energy, Office of Science, National Quantum Information Science Research Centers, Co-design Center for Quantum Advantage (C2QA) under contract No. DE-SC0012704.  The views and conclusions contained in this document are those of the authors and should not be interpreted as representing official policies, either expressed or implied, of the ARO or the US Government. The US Government is authorized to reproduce and distribute reprints for Government purposes, notwithstanding any copyright notation herein. Fabrication facilities use was supported by the Yale Institute for Nanoscience and Quantum Engineering (YINQE) and the Yale SEAS Cleanroom.
\end{acknowledgments}

\end{document}

%\preprint{APS/123-QED}

\title{%
  Supplementary Information \\
  \large ``Quantum Control of an Oscillator with a Kerr-cat Qubit''}

\author{Andy Z. Ding}
\thanks{These authors contributed equally to this work.\\ zhenghao.ding@yale.edu \\ benjamin.brock@yale.edu}
\author{Benjamin L. Brock}
\thanks{These authors contributed equally to this work.\\ zhenghao.ding@yale.edu \\ benjamin.brock@yale.edu}
\author{Alec Eickbusch}
\author{Akshay Koottandavida}
\author{Nicholas E. Frattini}
\author{\\  Rodrigo G. Corti\~{n}as}
\author{Vidul R. Joshi}
\author{Stijn J. de Graaf}
\author{Benjamin J. Chapman}
\author{Suhas Ganjam}
\author{Luigi Frunzio}
\author{Robert J. Schoelkopf}
\author{Michel H. Devoret}
\thanks{michel.devoret@yale.edu}
\affiliation{%
 Departments of Applied Physics and Physics, Yale University, New Haven, CT 06520, USA
}

\date{\today}
\maketitle
\onecolumngrid
\tableofcontents
\clearpage

\section{System Hamiltonian} \label{System Hamiltonian}

\subsection{Encoding a qubit in a Kerr-cat} \label{Encoding a qubit in a Kerr-cat}

In this section, we define the Bloch sphere of the KCQ \cite{Puri2017}, which is used throughout this work. The degenerate ground state manifold of the KCQ Hamiltonian $H_{\mathrm{KCQ}} = -Ka^{\dagger 2}a^2 + \epsilon_2 a^{\dagger 2} + \epsilon_2^* a^2$ is spanned by the orthornormal cat states $| \mathcal{C}^\pm_\alpha \rangle = \mathcal{N}_\alpha^\pm (|+\alpha \rangle \pm |-\alpha \rangle)$, where $a$ is the lowering operator of the SNAIL mode, $\alpha = \epsilon_2/K$ is the cat size and $\mathcal{N}_\alpha^\pm = 1/\sqrt{2\left( 1 \pm e^{-2\alpha^2}\right)}$ is the normalization coefficient. Without loss of generality, we take alpha to be real, $\alpha \in \mathbb{R}$. In the limit where $|\alpha|^2 \gg 1$, $\mathcal{N}_\alpha^\pm \rightarrow 1/\sqrt{2}$. We define the logical Pauli operators as%cardinal states on the Bloch sphere as
% \begin{equation}
%     \begin{split}
%         |\pm X \rangle & = | \mathcal{C}^\pm_\alpha \rangle \\
%         |\pm Y \rangle & = \left( | \mathcal{C}^+_\alpha \rangle \pm i| \mathcal{C}^-_\alpha \rangle \right) /\sqrt{2} = | \mathcal{C}^{\mp i}_\alpha \rangle \\
%         |\pm Z \rangle & = \left( | \mathcal{C}^+_\alpha \rangle \pm | \mathcal{C}^-_\alpha \rangle \right) /\sqrt{2},
%     \end{split}
% \end{equation}
% and the Pauli matrices as
\begin{equation}
    \begin{split}
        \sigma_x & = | \mathcal{C}^+_\alpha \rangle \langle \mathcal{C}^+_\alpha | - | \mathcal{C}^-_\alpha \rangle \langle \mathcal{C}^-_\alpha | \\
        \sigma_y & = i| \mathcal{C}^+_\alpha \rangle \langle \mathcal{C}^-_\alpha | - i| \mathcal{C}^-_\alpha \rangle \langle \mathcal{C}^+_\alpha | \\
        \sigma_z & = | \mathcal{C}^+_\alpha \rangle \langle \mathcal{C}^-_\alpha | + | \mathcal{C}^-_\alpha \rangle \langle \mathcal{C}^+_\alpha |
    \end{split}
\end{equation}
We can then define the projector $\mathcal{P}_\mathcal{C} = | \mathcal{C}^+_\alpha \rangle \langle \mathcal{C}^+_\alpha | + | \mathcal{C}^-_\alpha \rangle \langle \mathcal{C}^-_\alpha |$ to project the bosonic operators of the SNAIL $a$, $a^\dagger$ onto the KCQ subspace. 

\subsection{Coupling a KCQ to a cavity} \label{Coupling a KCQ to a cavity}

In this section we derive the static effective Hamiltonian of our setup consisting of a KCQ coupled to a storage cavity. Our calculation follows the method in \cite{Venkatraman2022, Xiao2023} where we use Feynman-like diagrams to systematically compute the static terms describing different cascaded nonlinear processes in our setup. \\

The Hamiltonian of the system consisting of a SNAIL coupled to a cavity is given by

\begin{equation}
    H_\mathrm{full} = H_\mathrm{SNAIL} + H_\mathrm{cavity} + H_c + H_d
    \label{system_H}
\end{equation}
where $H_\mathrm{SNAIL} = \omega_a' a'^{\dagger} a' + \sum_{n>2} g_n (a' + a'^\dagger)^n$ is the SNAIL Hamiltonian, $H_\mathrm{cavity} = \omega_b' b'^\dagger b'$ is the Hamiltonian of the microwave cavity, $H_c = g (a' + a'^\dagger) (b' + b'^\dagger)$ is the capacitive coupling between the two modes, $H_d = 2 \Re(e^{i\omega_s t}) (\epsilon_s a'^\dagger + \epsilon^\ast_s a') + 2 \Re(e^{i\omega_d t}) (\epsilon_d a'^\dagger + \epsilon^\ast_d a')$ is the drive Hamiltonian, $\epsilon_s$ is the strength of the squeezing drive on the SNAIL, and $\epsilon_d$ is the resonant drive on the SNAIL. In these expressions, $a'$ and $b'$ are the bare modes of the SNAIL and the cavity, with bare frequencies $\omega_a'$ and $\omega_b'$, respectively. The capacitive coupling hybridizes the modes and shifts their frequencies. Therefore, we reserve the notations $a$ and $b$ for the hybridized modes that we can measure experimentally. \\

We truncate the SNAIL nonlinearity to the fourth order to compute the static effective Hamiltonian of our system,

\begin{equation}
    H = \Delta_{s} a^\dagger a - K a^{\dagger 2}a^2 + \epsilon_2 a^{\dagger 2} + \epsilon_2 a^2 - K_b b^{\dagger 2}b^2  - \chi_{ab}a^\dagger a b^\dagger b 
    \label{Eq.kerr-cat_cavity}
\end{equation}
where we have, to the leading order of $(g/\Delta)^2$ or the leading non-negligible terms,

\begin{equation}
    \begin{split}
        \Delta_s & = \left( \frac{24g_4}{9g_3^2} - \frac{9}{\omega_a} \right) \epsilon_2^2 \\
        K & = -6 g_4 + \frac{30g_3^2}{\omega_a} \\
        K_b & = -6 g_4 \left( \frac{g}{\Delta} \right)^4 + 9g_3^2\left( \frac{g}{\Delta} \right)^2 \left( \frac{1}{2\omega_a - \omega_b} - \frac{1}{2\omega_a + \omega_b} - \frac{4}{\omega_b} \right) \\
        \chi_{ab} & = -24 g_4 \left( \frac{g}{\Delta} \right)^2 + 36g_3^2\left( \frac{g}{\Delta} \right)^2 \left( \frac{1}{2\omega_a - \omega_b} + \frac{1}{2\omega_a + \omega_b} + \frac{2}{\omega_a} \right).
    \end{split}
\end{equation}
Plugging in the parameters from Table. \ref{Table of parameters}, we calculate $(g/\Delta)^2 = 0.5\%$, so we can safely ignore higher order terms in the expansion of $\Delta_s$, $K$, and $\chi_{ab}$. In the case of $K_b$ the leading order of the expansion is already of order $(g/\Delta)^4$. Given that $K_a/2\pi = 0.93$ MHz, the inherited self-Kerr of the storage from the SNAIL is negligible, which makes this setup suitable for encodings such as GKP states \cite{GKP2001_sp} that are susceptible to Kerr effects.\\

To gain intuition for the dispersive coupling between a KCQ and a storage, it is useful to consider the projection of the SNAIL number operator $a^\dagger a$ onto the KCQ subspace
\begin{equation}
    \begin{split}
        \mathcal{P}_\mathcal{C} a^\dagger a \mathcal{P}_\mathcal{C} & = \alpha^2 r^2 | \mathcal{C}^+_\alpha \rangle \langle \mathcal{C}^+_\alpha | + \alpha^2 r^{-2} | \mathcal{C}^-_\alpha \rangle \langle \mathcal{C}^-_\alpha | \\
        & = \alpha^2 \left( \mathbb{I} - 2e^{-2\alpha^2} \sigma_x \right). \\
    \end{split}
\end{equation}
After projection, the number operator of the SNAIL mode gets turned into an identity and a $\sigma_x$ rotation. Of particular interest to us is the scaling factor in front of the $\sigma_x$ rotation term. As we increase the squeezing amplitude, this rotation gets exponentially suppressed. \\

\subsection{KCQ induced cavity dephasing} \label{KCQ induced cavity dephasing} 

\begin{figure}
\includegraphics[width=0.99\textwidth]{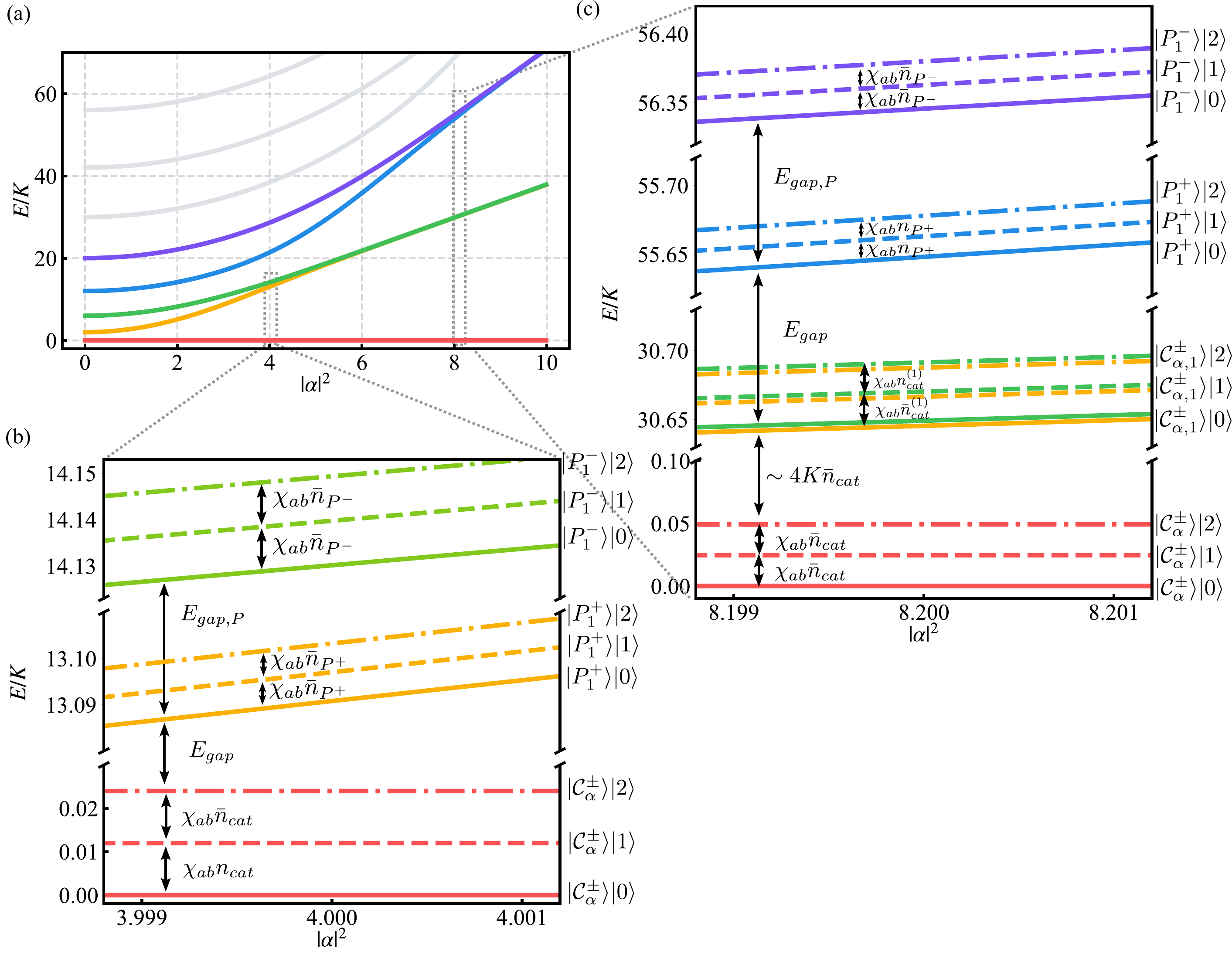}
\caption{Spectrum of a KCQ coupled to a cavity as a function of cat size $\alpha$, obtained from numerical diagonalization of the system Hamiltonian in Eq. \ref{Eq.kerr-cat_cavity}.}\label{Fig. diagonalization}
\end{figure}

To better understand the interaction between a KCQ and a storage cavity, we diagonalize the Hamiltonian in Eq. \ref{Eq.kerr-cat_cavity} using our experimental parameters in Table \ref{Table of parameters}. The spectrum we obtain from the diagonalization as a function of $|\alpha|^2$ is shown in Fig. S \ref{Fig. diagonalization}. At an energy scale comparable with $K$ in Fig. S \ref{Fig. diagonalization}(a), the spectrum agrees with the result from \cite{Frattini2022}, where the excited states of the KCQ become degenerate as we increase $|\alpha|^2$. The gaps between different lines, such as $E_{\mathrm{gap}}$ and $E_{\mathrm{gap, P}}$ are determined by the drive strength and $K$ of the KCQ, where the subscript $P$ indicates that the states have definite photon number parity. When we zoom in on the lines at different cat sizes in Fig.\ref{Fig. diagonalization}(b) and (c), we see the spectrum of the KCQ split into different harmonic sectors due to coupling with Fock states in the storage cavity. The frequency of each harmonic sector is a function of the average photon number of the KCQ state. \\

As an example, in the case of Fig. S \ref{Fig. diagonalization}(b) where $|\alpha|^2=4$, only the degenerate ground states of the KCQ, spanned by $|\mathcal{C}_\alpha^\pm \rangle$ are inside the double well potential. The storage photons split the ground states evenly by $\chi_{ab}\bar{n}_{\mathrm{cat}}$, where $\bar{n}_{\mathrm{cat}}=|\alpha|^2$. This indicates that the storage photons are not able to distinguish between the different cat states in the KCQ ground state manifold, the storage is simply Stark-shifted by $\chi_{ab}\bar{n}_{\mathrm{cat}}$. Therefore, the storage cavity and the KCQ will not entangle with each other during idling time when the KCQ stays in its ground state. The first excited states are parity states right above the wells. Though the storage photons still split each level evenly, different levels in the KCQ Stark-shift the storage cavity by different amounts due to their different average photon numbers $\bar{n}_{P^-} \neq \bar{n}_{P^+} \neq \bar{n}_{\mathrm{cat}}$. As a result, the storage cavity is capable of distinguishing between different levels in the KCQ, since it rotates at slightly different frequencies depending on whether the KCQ is in its ground or excited states. \\

\begin{figure}
\includegraphics[width=0.95\textwidth]{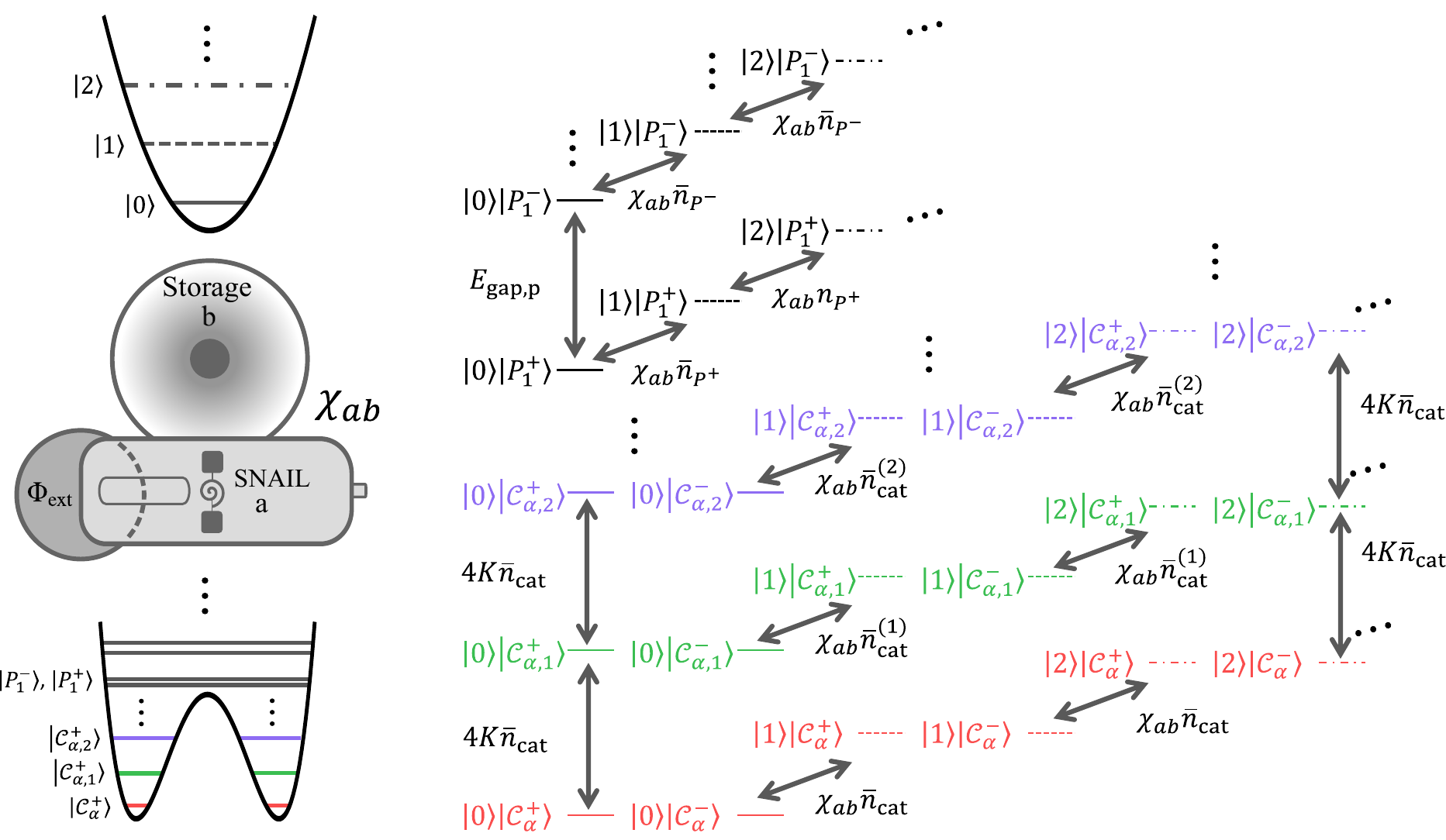}
\caption{Energy level diagram of a system with a KCQ coupled to a storage cavity. }\label{Fig.energy level diagram}
\end{figure}

This understanding can be extended to the case of larger KCQs with more than one level inside the wells, as in Fig. S \ref{Fig. diagonalization}(c) with $|\alpha|^2=8.2$. In this case, the spectral lines of the first excited states are merging while the second excited states are still outside of the wells. In this case, the Stark shift of the storage cavity when the KCQ is in its first excited state is still distinct from that of the ground state of the KCQ, as we have $\bar{n}_{\mathrm{cat}}^{(1)} \approx \bar{n}_{\mathrm{cat}} + 1$. In the limit of large $\alpha$, we have two distinct regions: states inside the well and states outside the well. Fig. S \ref{Fig.energy level diagram} shows the energy level diagram in this asymptotic limit. \\

In the absence of KCQ heating, the cavity coherence is protected from dephasing at large $\alpha$ because the cross-Kerr interaction simplifies into a Stark-shift of the storage cavity, which can be tracked in software. When the KCQ is in an excited state, however, the storage cavity is Stark-shifted by a different amount. Thus, when the KCQ heats to an excited state for an unknown amount of time, the storage accumulates an unknown amount of phase, leading to dephasing. This dephasing is what we have measured in Fig. 3(c) of the main text. \\

To model the cavity coherence in the absence of KCQ heating, as well as the suppression of cavity dephasing at small $\epsilon_2$, we perform a master equation simulation using Eq. \ref{Eq.kerr-cat_cavity} in the previous section as our model of the system and photon loss as the only dissipation. We manage to achieve good agreement between data and simulation for the regime $|\alpha|^2 < 1$ only by setting $T_{1,a} = 100$ $\mu s$ for the SNAIL, which is significantly different from our characterized SNAIL coherence in Table \ref{Table of parameters}. This is due to a lack of understanding of the heating mechanism of the KCQ \cite{Frattini2022, Venkatraman_note, Hajr2024}. As a strongly driven nonlinear quantum system, the environment seen by the KCQ is not Markovian, rendering our master equation simulation only a preliminary attempt at understanding the dynamics of the system. As a result, we did not push the limit of the simulation to match our data by adding in dephasing or SNAIL heating. This is also because we are using the simulation result as a ``guide-to-the-eye" for our data on cavity coherence under active cooling on the KCQ, showcasing that in the absence of KCQ heating, the cavity dephasing will be suppressed when we increase the squeezing amplitude. \\

\section{System setup} \label{System setup}

\subsection{Device fabrication} \label{Sec. Device fabrication}

Following the recipe in \cite{Ganjam2024}, the on-chip superconducting circuits used in our experiment were fabricated on a c-plane sapphire substrate grown using the heat-exchange method (HEM). The substrate wafer was cleaned using a piranha solution (2:1 $\mathrm{H_2SO_4:H_2O_2}$) for 20-25 minutes and rinsed under running deionized (DI) water before being annealed in a FirstNano EasyTube 6000 furnace. The annealing step has been shown to improve substrate surface and bulk loss \cite{Ganjam2024}. During the annealing process, the furnace was preheated to 200\degree C with constant oxygen flow before loading the sample. It was then heated up to 1200\degree C at a ramp rate of 400\degree C per hour. Afterwards, the furnace was held at 1200\degree C for 1 hour before cooled down gradually to room temperature over 6 hours with constantly flowing gas mixture (4:1 $\mathrm{N_2:O_2}$). \\

The tantalum layer was uniformly sputtered on the substrate using the Kurt J. Lesker CMS 18 DC magnetron sputtering system. The substrate was loaded into the sputtering chamber with a niobium thermalization disc in the back for better heating uniformity. Afterward, the substrate was heated to 400\degree C and was held for 15 minutes for dehydration bakeout. The substrate was then heated to the deposition temperature at 800\degree C and idled for 10 minutes for temperature uniformity. The tantalum was sputtered at 300 W with 6 mTorr of Ar gas at a rate of 2.5\AA/s for 10 minutes. The system then idled again for 10 minutes at 800\degree C for the post-deposition bake. Subsequently, the system was cooled down at a controlled rate of 10\degree C per minute down to 500\degree C. This was to release the stress in the tantalum film into the substrate adiabatically to avoid material damage at the metal-substrate interface. Afterward, the substrate was cooled down to room temperature at a faster rate. It took 3-5 hours to reach room temperature before the substrate was unloaded.\\

Before optical lithography, the substrate surface was primed with HMDS vapor for better resist adhesion and uniformity. During the process, the substrate was loaded into a TA Series Yield Engineering System (YES) oven right after the annealing process. The oven chamber was then purged with HMDS vapor three times before baking out the wafer at 150\degree C for 25 minutes. At the end of the process, the chamber was re-purged with HMDS vapor three times before the substrate was removed.\\

The sapphire substrate was then coated with SC1805 photoresist using Laurell spin coater WS-400 at 500 rpm for 10 seconds before ramping up to 4000 rpm for 1 minute and 30 seconds for film uniformity. The resist was baked at 115\degree C for 1 minute. The edge beads were removed using acetone before the wafer was loaded into the Heidelberg Maskless Aligner 150 direct laser writer for exposure with a calibrated dose of 92 mJ/mm$^2$. Then the resist was developed with Microposit MF319 developer for 1 minute and then treated with oxygen plasma using the AutoGlow 200 plasma system for 1 minute and 30 seconds at 150 W and 300 mTorr of oxygen. \\

The tantalum layer was etched using the dry etch recipe from \cite{Ganjam2024} with an Oxford 80 reactive ion etch system. Before the sample was loaded, the chamber was prepped using the standard cleaning recipe, which consists of two main steps. We ran the recipe with an empty chamber before the actual etching. The first step is 10 minutes of cleaning with a mixture of O$_2$ and SF$_6$ at flow rates 50 sccm and 10 sccm, respectively. The power was at 100 W and the pressure was maintained at 50 mTorr. The second step is another 10 minutes of cleaning with a mixture of O$_2$ and Ar at 50 sccm and 10 sccm, respectively. The power was kept at 50 W and the pressure was 50 mTorr. These two steps conditioned the chamber to have more reproducible results. The tantalum etch recipe required us to flow SF$_6$ at a rate of 20 sccm at a pressure of 10 mTorr for 3 minutes to etch through 150 nm of tantalum. After etching, the substrate was cleaned with 3 minutes of sonication in N-Methylpyrrolidone (NMP), acetone and isopropyl alcohol (IPA) sequentially to get rid of the photoresist and was blown dry with nitrogen. \\

We thoroughly cleaned the wafer before electron beam (E-beam) lithography to reduce surface loss. The substrate was cleaned with a piranha solution (2:1 $\mathrm{H_2SO_4:H_2O_2}$) for 20-25 minutes to get rid of organic contaminants. Then it was rinsed under DI water for 5-10 minutes before being transferred in Transene 10:1 buffered oxide etch (BOE) for 20 minutes to etch away most of the tantalum oxides \cite{McLellan2023}. The DI water rinse prior to BOE is crucial because a mixture of piranha and BOE solutions can attack tantalum, a phenomenon we learned through trial and error. The BOE step was followed by a 20-25 minute rinse in DI water.\\ 

The SNAIL loops and the Josephson junctions were patterned using electron-beam lithography. After a dehydration bake at 180\degree C for 5 minutes, the substrate was spin-coated with ~800 nm of MMA (8.5) MAA EL13 and ~200 nm of 950K PMMA A4, with a 5-minute bake at 180\degree C after each layer was spun. Then, a 15 nm aluminum anticharging layer was deposited on top of the bilayer to avoid charging effects. The bridge-free junctions were then written using a Raith EBPG 5200+ at a base dose of 290 $\mu$C/cm$^2$. Afterwards, the anticharging layer was removed using MF312 developer for ~2 minutes. The Ebeam patterns were developed in 3:1 IPA:H$_2$O for 2 minutes at 6\degree C and the wafer was blown dry with nitrogen. \\

The Josephson junctions were deposited using a Plassys UMS300 electron-beam evaporator. After being loaded in the load lock and pumped to base pressure, the sample was etched with an Ar ion beam at 400 V and 22 mA at $\pm 45\degree$ angles for 34 seconds each. This step removed the oxide on the tantalum sidewalls and cleaned the substrate regions where aluminum was to be deposited. Subsequently, the aluminum was evaporated at angles of $\pm 20\degree$ in the evaporation chamber with an oxidation step in the middle using an 85:15 Ar:O$_2$ mixture at 15 Torr for 15 minutes. After the depositions, the surface of the devices was capped with a layer of aluminum oxide grown at 50 Torr for 5 minutes with the same gas mixture in the oxidation chamber before the substrate was unloaded. \\

The liftoff was performed by submerging the substrate in NMP at 75\degree C for 4 hours. Then it was cleaned by being sonicated in RT NMP, acetone, and IPA for 1 minute each. Then the wafer was blown dry with nitrogen. A layer of SC1827 photoresist was coated on the wafer before it was diced with an ADT ProVectus 7100 dicer. The chips were cleaned by being sonicated in NMP, acetone, and IPA for 2 minutes each and were tested by probing at room temperature before being loaded into our experimental package.\\

\subsection{Package processing}

The main 3D aluminum package was mechanically machined out of high-purity (5N5) aluminum. It was then cleaned by sonicating in NMP, acetone and IPA for 5 minutes each. Afterward, it went through a chemical etching treatment using Transene Aluminum Etchant Type A to get rid of contaminants from the machining process \cite{Reagor2013}. During the etching process, the package was mounted on a custom-designed PTFE cage and submerged in 1L of the etchant solution that was heated to ~50\degree C for about 2 hours in total. During the etching process, the etchant bath needs to be periodically changed to avoid secondary reaction product redepositing on the surface of the package. Also, the bath was stirred at 600 RPM. Afterwards, the package was rinsed thoroughly with DI water for 10 minutes, before blown dry with nitrogen gas. The auxiliary parts, including the chip clamps, the copper brackets and the screws were cleaned by sonication in NMP, acetone and IPA for 3 minutes each.\\

\subsection{Wiring diagram} \label{Wiring diagram}

\begin{figure}
\begin{center}
\includegraphics[width=0.8\textwidth]{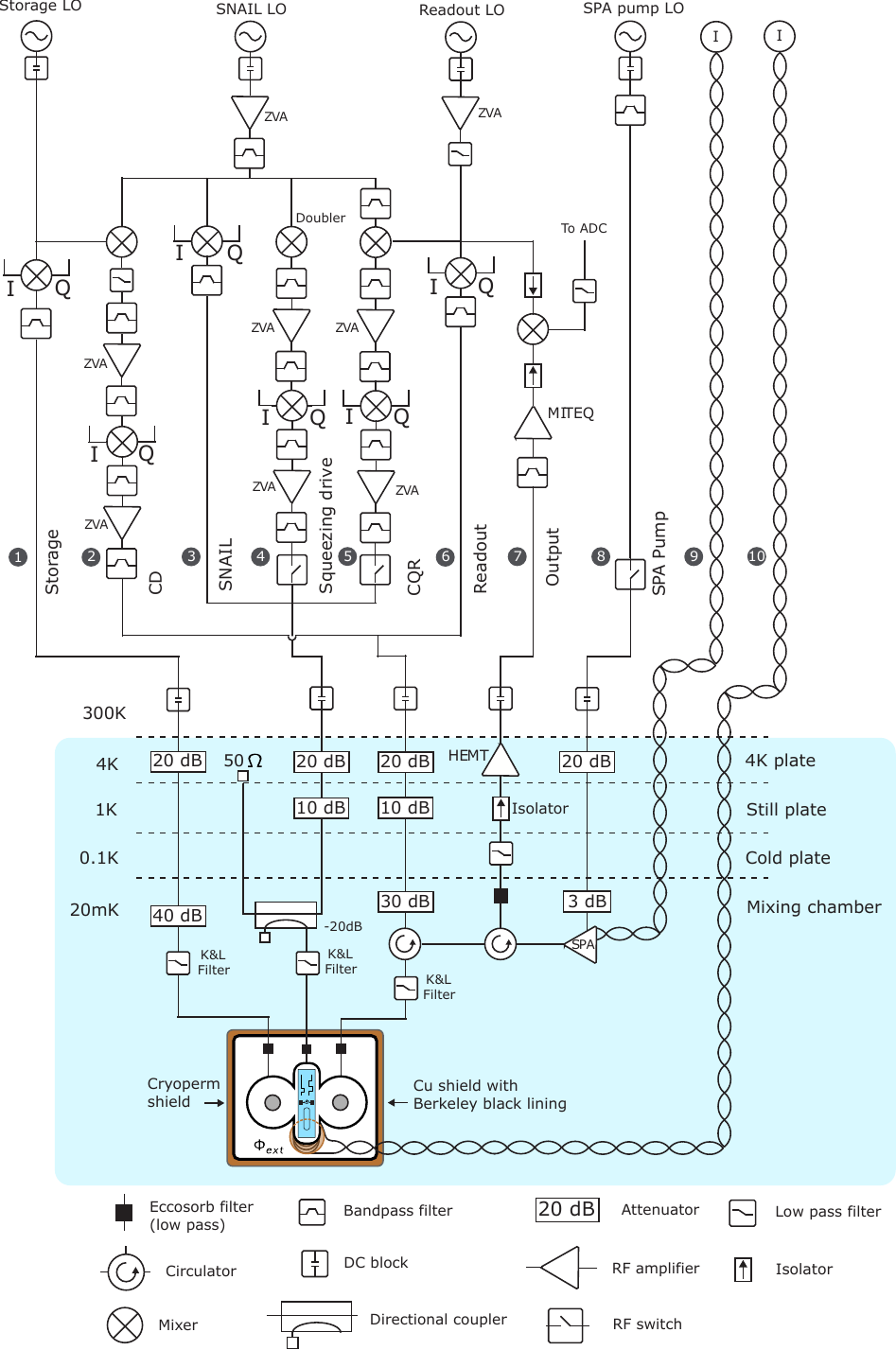}
\caption{The wiring diagram of our setup.}\label{Fig.Wiring diagram}
\end{center}
\end{figure}

Our experimental device is clamped to the mixing chamber of a dilution refrigerator, as depicted in Fig. S\ref{Fig.Wiring diagram}, and surrounded with two layers of shielding.  The outermost shield is a cryoperm $\mathrm{\mu}$-metal magnetic shield, and the innermost shield is made of copper and coated with Berkeley black \cite{Persky1999} to absorb stray infrared radiation. To characterize and control our experimental device inside the dilution refrigerator, we built a set of microwave lines shown in the wiring diagram in Fig. S. \ref{Fig.Wiring diagram}. There are 10 lines in total, and each line is numbered next to its respective label on the diagram. \\

Lines 1, 3, and 6 provide the microwave signals for resonantly driving the storage cavity, SNAIL mode, and readout cavity, respectively. On these lines a signal generator supplies a local oscillator (LO) tone to an IQ mixer, which up- or down-converts the LO tone via the inputs to its in-phase (I) and quadrature (Q) ports.  These IQ pairs are sourced from an FPGA-based digital-to-analog converter (DAC) with custom software \cite{Ofek2016}, providing us full control of the amplitude and phase of our drives with $2$ ns sampling resolution.\\

Lines 2, 4, and 5 provide the microwave signals for parametrically driving the conditional displacement (CD), squeezing, and cat-qubit readout (CQR) processes, respectively. On line 2 we mix together the SNAIL LO and storage LO (using), filter the output to isolate the desired difference frequency, and amplify this tone (with a Mini-Circuits ZVA-183-S+ amplifier) so it can provide sufficient power to the LO input of an IQ mixer. This tone is then mixed with an IQ DAC pair from our FPGA to give us full control of the amplitude and phase of our CD drive, and then amplified to achieve sufficient power to drive this parametric process. Line 4 is designed in the same way, except we are mixing the SNAIL LO with itself (using a Marki MLD-0416LSM frequency doubler) to generate a tone at twice the frequency of this signal. Line 5 is designed in the same way, except we are mixing together the SNAIL LO with the readout LO (using a Marki M2H-0220HP mixer) to generate a tone at the difference frequency between these signals. \\

\begin{figure}
\includegraphics[width=1.0\textwidth]{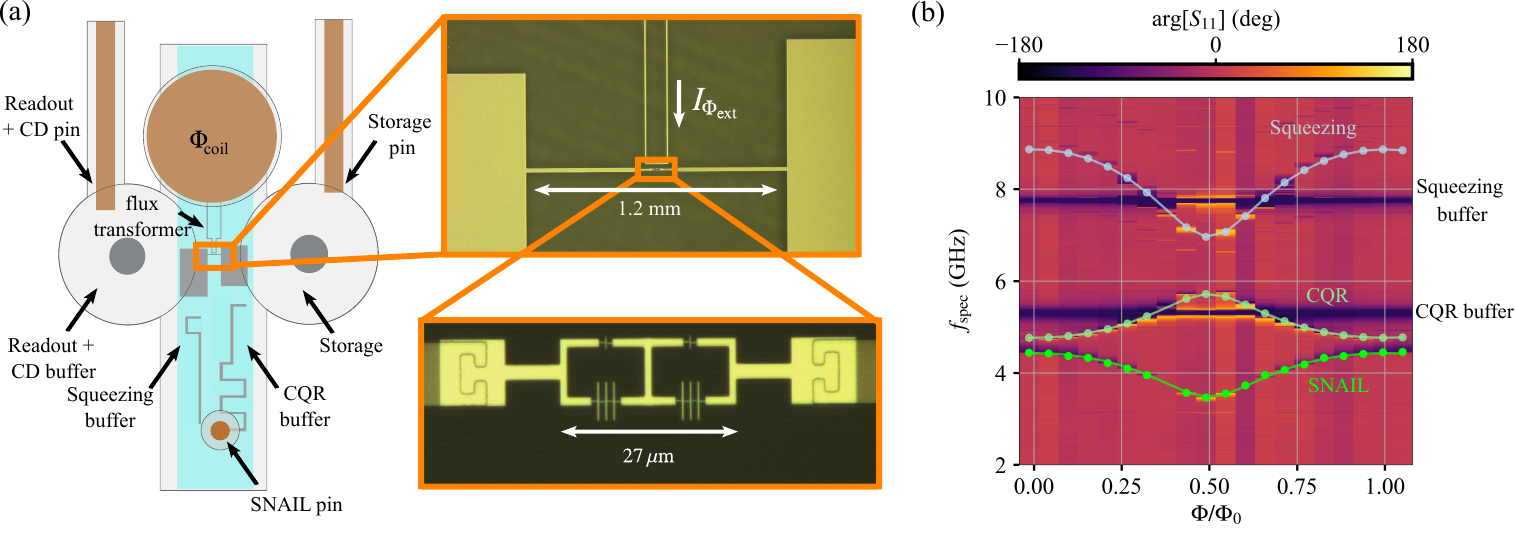}
\caption{\textbf{The setup and basic spectroscopy result.} (a) A cartoon of our experimental setup modeled from above. The sapphire chip (colored in blue) with the superconducting circuits (colored in silver) sits in between two 3D aluminum post cavities (colored in silver) with coupling pins (colored in brown) for delivering microwave control lines. The SNAIL has its drive pin perpendicular to the chip at the bottom end of the chip tunnel. Above the chip, there is a copper coil (colored brown) wound with Nb-Ti wires for threading magnetic flux into our SNAIL loops. On the right side of (a), we have the optical micrograph of our SNAIL circuit, where the SNAIL loops with the Josephson junctions are made of aluminum, and the rest of the structures are made out of tantalum. (b) The two-tone spectroscopy result, where we drive the coupling pin to the SNAIL with a CW spectroscopy tone while measuring the reflection coefficient of the readout cavity. Whenever photons are absorbed by a mode the readout is Stark-shifted, which manifests as a change in the phase of the reflection coefficient S11. The green curve is our best fit for the SNAIL frequency as a function of flux. The light green curve is the difference between the CQR buffer mode frequency and the fitted SNAIL frequency, corresponding to the CQR parametric process. The light blue curve is twice the fitted SNAIL frequency, corresponding to the squeezing parametric processes.}\label{Sp1}
\end{figure}

After conditioning all of these drive signals, they are combined in different ways and sent into the dilution refrigerator. The signal on line 1 is delivered to the storage pin (see Fig. S\ref{Sp1}(a) for our nomenclature regarding the pins). After entering the fridge it is attenuated (nominally to thermalize noise in the transmission line to the temperature of the mixing chamber) and filtered (with a K\&L 6L250-12000/T26000-O/O tubular low-pass filter and an eccosorb filter) before arriving at the storage pin. The eccosorb filter is used to attenuate infrared Cooper-pair-breaking radiation. The signals on lines 2 and 6 are combined and delivered to the readout pin. After entering the fridge this combined signal is similarly attenuated and filtered before arriving at the readout pin, but in this case it also passes through a cryogenic circulator (Quinstar OXE89). The signals on lines 3, 4, and 5 are combined and delivered to the SNAIL pin. After entering the fridge this combined signal is similarly attenuated and filtered before arriving at the SNAIL pin, but in this case we use a $20$ dB directional coupler instead of an attenuator at the mixing chamber. This syphons off $1\%$ of the signal and sends the remaining $99\%$ back up the fridge to the $4$K plate to be dissipated as heat. In this way we are able to decouple the process of attenuation from heat generation; if we were to use a 20 dB attenuator instead, the fridge would heat up due to the high power needed for the parametric drives being delivered to the SNAIL pin.\\

Line 7 is our output line. Photons in the readout cavity that leak out through the readout pin are first amplified by our quantum-limited SNAIL parametric amplifier (SPA) \cite{Frattini2017}, then a cryogenic HEMT amplifier, and finally a room-temperature amplifier (MITEQ AMF-5F-04001200). The resulting signal is then mixed together with the readout LO tone (using a Marki IRW-0618LXW-2 image reject mixer), resulting in $50$ MHz signal that is digitially acquired using the analog-to-digital (ADC) of our FPGA. \\

Lines 8 and 9 are used for operating our SPA.  Line 8 simply provides a pump tone to the SPA, while line 9 delivers a DC current that threads flux through the SPA to tune its operating point. This DC current is carried by a twisted pair of superconducting NbTi wires from the $4$K plate down to the mixing chamber of our dilution refrigerator. To tune the SPA we only need to drive a current of a few hundred $\mu$A, so we are able to modularize this DC line using normal-metal connectors at both the $4$K plate and the mixing chamber.\\

Line 10 delivers a DC current that threads flux through our on-chip SNAIL, enabling us to tune its operating point. The DC current is again carried by a twisted pair of superconducting NbTi wires from the $4$K plate down to the mixing chamber, but in this case we need to drive about $100$ mA of current to tune the SNAIL to its operating flux bias point of $0.32\Phi_0$. If we were to use normal-metal connectors anywhere below the $4$K plate, the heat generated would overwhelm the cooling power of the fridge. To get around this, we use a continuous NbTi twisted pair running from $4$K to the mixing chamber, following the method introduced in \cite{Chapman2023}.

\section{System design and characterization} \label{System basics and calibration}

\subsection{Design and simulation}

Our experimental package consists of a sapphire chip (50 mm $\times$ 5.6 mm) sandwiched between two 3D aluminum post cavities, as is depicted in Fig. S \ref{Sp1}(a). Superconducting circuits are fabricated on the chip out of tantalum and aluminum, as described above. The package was designed using electromagnetic finite element solver HFSS. The SNAIL circuit is at the center of the chip, highlighted in the orange boxes. The SNAIL consists of two superconducting loops to lower its anharmonicity \cite{Frattini2017, FrattiniThesis}. As the SNAIL circuit is a flux-tunable device, we need to deliver external magnetic flux to our device inside our experimental setup. To do this, we follow the experimental setup developed in \cite{MundhadaThesis, Chapman2023}, using a copper coil nested inside the aluminum package as well as an on-chip flux transformer. When we run DC current in the twisted pair, the current will turn the copper coil into an electromagnet, which then threads flux into the flux transformer loop on the sapphire chip underneath. The flux generates current in the loop that travels down the chip close to the SNAIL loops, where its electromagnetic field threads flux into the SNAIL circuit, providing it with the external DC flux bias. In our design process, we optimized the geometry of the pick-up loop of the flux transformer so that its eigenmodes do not crowd into frequencies of interest in the package. The common mode of the flux transformer is at $\omega_c/2\pi = 9.26$ GHz, and its differential mode is at 9.76 GHz. The common mode has a cross-Kerr $\chi_{ac}/2\pi=2.31 \pm 0.1$ KHz to the SNAIL mode with a total linewidth of $\kappa_c/2\pi = 131.4 \pm 0.4$ kHz. Compared with the readout mode, it enjoys a much larger $\chi/\kappa$ ratio and is therefore more suitable for dispersive readout. As a result, in our basic system characterization experiments, we use the common mode of the flux transformer as our readout mode to perform two-tone spectroscopy, which is discussed in more detail in the next subsection. \\

The SNAIL circuit hosts the KCQ, and it is driven with microwave signals delivered via the SNAIL pin shown at the bottom end of the chip tunnel in Fig. S \ref{Sp1}. The SNAIL pin is coupled to the SNAIL at a rate of $(150$ $\mu$s$)^{-1}$, which corresponds to the Purcell limit of this mode. Inspired by the design in \cite{Chapman2023}, we design two on-chip meandering resonators serving as buffer modes for the squeezing drive (on the left at 7.760 GHz with a linewidth of $1.35 \pm 0.02$ MHz) and the CQR drive (on the right at 5.308 GHz with a linewidth of $1.24 \pm 0.01$ MHz). The squeezing drive at 7.996 GHz is detuned from the squeezing buffer by 240 MHz, and the CQR drive at 5.362 GHz is detuned from its buffer mode by 56 MHz. Both parametric drives are very far away from their designed buffer modes in frequency. Nevertheless, we can still efficiently drive these parametric processes without using excessively noisy amplifiers (see the wiring diagram in Fig. S \ref{Fig.Wiring diagram}): we use Minicircuits ZVAs instead of ZVEs, as were used in \cite{Grimm2020, Frattini2022, Venkatraman2022}. The capacitor pads of the SNAIL circuit have a slight asymmetry to avoid coupling the two post cavities directly via the common mode of the circuit. The post cavity on the right (7.025 GHz) is our high-Q storage cavity. The post cavity on the left serves two purposes. Its fundamental mode (3.133 GHz, with a linewidth of $0.28 \pm 0.01$ MHz in total) is the buffer mode for the CD parametric drive, and its first harmonic (9.362 GHz) is the readout cavity for the SNAIL/KCQ. The CD drive at 3.027 GHz is detuned from its buffer by 106 MHz, but we can still drive $g_{CD}/2\pi = 3.1$ MHz (see CD calibration below). \\

\subsection{System characterization with two-tone spectroscopy} \label{System calibration with two-tone spectroscopy}

To experimentally validate the microwave design of our package, as well as extract the SNAIL parameters, we perform a two-tone spectroscopy experiment as a function of the SNAIL’s flux bias. In this experiment, we drive the SNAIL pin with a CW spectroscopy tone swept from 2-10 GHz at -50 dBm while measuring the reflection coefficient of the cavity via the readout + CD pin. Whenever photons are absorbed by a mode, the readout is Stark-shifted, which manifests as a change in the phase of the reflection coefficient S$_{11}$. In the results shown in Fig. S \ref{Sp1}(b), we identify the resonance feature tuning with flux from 3.5 - 4.5 GHz as our SNAIL mode, the feature tuning from 4.9-5.9 GHz as our CQR parametric process, and the feature tuning from 6.9-8.9 GHz as our squeezing parametric process. We extract the SNAIL parameters listed in the top section of Table \ref{Table of parameters} by fitting the measured SNAIL frequencies using the model $H_\mathrm{SNAIL} = \omega_a' a'^{\dagger} a' + \sum_{n>2} g_n (a' + a'^\dagger)^n$ up to $n=4$. We also identify other prominent features in the two-tone spectroscopy result, such as the CQR parametric process and the squeezing process. We subtract the fitted SNAIL frequencies from the CQR buffer frequency and the squeezing buffer frequency to obtain the lines in light green and light blue. They serve as guide-to-the-eye for the measured CQR and squeezing processes. The readout mode and the CD buffer mode are barely visible as they are not as strongly coupled to the SNAIL drive pin by design.\\

\begin{table}
\begin{tabular}[c]{ l|c|l } 
\hline
\hline
\textbf{Parameter} & Value & Measurement or estimate method \\
\hline
SNAIL charging energy $E_\mathrm{C}/h $& 62 MHz & Two-tone spectroscopy and SNAIL fit \\ 
Number of SNAILs & 2 & Design \\ 
SNAIL asymmetry $\alpha_{\mathrm{SNAIL}}$ & 0.11 & Two-tone spectroscopy and SNAIL fit \\
SNAIL 1 large junction inductance & 0.6 nH & Two-tone spectroscopy and SNAIL fit \\
SNAIL 1 small junction inductance & 5.5 nH & Two-tone spectroscopy and SNAIL fit \\
SNAIL linear stray inductance $L_{\mathrm{lin}}$ & 1.5 nH & Two-tone spectroscopy and SNAIL fit \\
SNAIL frequency at $\Phi/\Phi_0=0$ & 4.46 GHz & Two-tone spectroscopy \\
SNAIL frequency at $\Phi/\Phi_0=0.5$ & 3.46 GHz & Two-tone spectroscopy \\
\hline
SNAIL operating bias point $\Phi_\mathrm{op}/\Phi_0$ & 0.32 & Design \\
SNAIL operating frequency $\omega_a/2\pi$ & 3.998 GHz & Two-tone spectroscopy \\
SNAIL cubic nonlinearity $g_3/2\pi$ & 6 MHz & Two-tone spectroscopy and SNAIL fit \\
SNAIL self-Kerr nonlinearity $K/2\pi$ & 0.93 $\pm$ 0.03 MHz & Kerr refocusing msmt.  \\
SNAIL single-photon decay time $T_{1,a}$ & 16.0 $\pm$ 0.4 $\mu$s & Coherence msmt. \\
SNAIL Ramsey decay time $T_{2,a}$ & 7.2 $\pm$ 0.1 $\mu$s & Coherence msmt. \\
SNAIL Hahn echo decay time $T_{2\mathrm{E},a}$ & 9.5 $\pm$ 0.1 $\mu$s & Coherence msmt. \\
\hline
Storage cavity frequency $\omega_b/2\pi$ & 7.025 GHz & Two-tone spectroscopy \\
Storage cavity frequency tuning range with external flux & 1.2 MHz & Two-tone spectroscopy \\
Storage cavity single-photon decay time $T_1$ & 204 $\pm$ 9 $\mu$s & Coherence msmt. \\
Storage cavity Ramsey decay time $T_2$ & 381 $\pm$ 8 $\mu$s & Coherence msmt. \\
Storage cavity to SNAIL cross-Kerr $\chi_{ab}/2\pi$ & 2.91 $\pm$ 0.03 kHz & Coherence msmt. \\
Storage cavity self-Kerr nonlinearity $K_b/2\pi$ & $<1$ Hz & Simulation \\
\hline
Readout cavity frequency $\omega_r/2\pi$ & 9.362 GHz & Direct RF reflection msmt. \\
Readout cavity total linewidth $\kappa_r/2\pi$ & 0.396 MHz & Direct RF reflection msmt. \\
Readout cavity internal linewidth & 0.04 MHz & Direct RF reflection msmt. \\
Readout cavity to SNAIL cross-Kerr $\chi_{ar}/2\pi$ & 1.51 $\pm$ 0.01 kHz & Direct RF reflection msmt. \\ 
\hline
CQR buffer mode frequency & 5.308 GHz & Two-tone spectroscopy \\
CQR buffer mode linewidth & 1.24 $\pm$ 0.01 MHz & Two-tone spectroscopy \\
Squeezing buffer mode frequency & 7.760 GHz & Two-tone spectroscopy \\
Squeezing buffer mode linewidth & 1.35 $\pm$ 0.02 MHz & Two-tone spectroscopy \\
CD buffer mode frequency & 3.133 GHz & Two-tone spectroscopy \\
CD buffer mode linewidth & 0.28 $\pm$ 0.01 MHz & Two-tone spectroscopy \\
\hline
\hline
\end{tabular}
\caption{\textbf{Summary of device parameters.} The design simulations are carried out with Ansys HFSS and the black-box quantization method \cite{FrattiniThesis, Nigg2012, Minev2021}.}
\label{Table of parameters}
\end{table}

\section{KCQ calibration and characterization} \label{KCQ calibration and characterization}

\subsection{Calibrating CQR}
\begin{figure}[h!]
\includegraphics[width=0.4\textwidth]{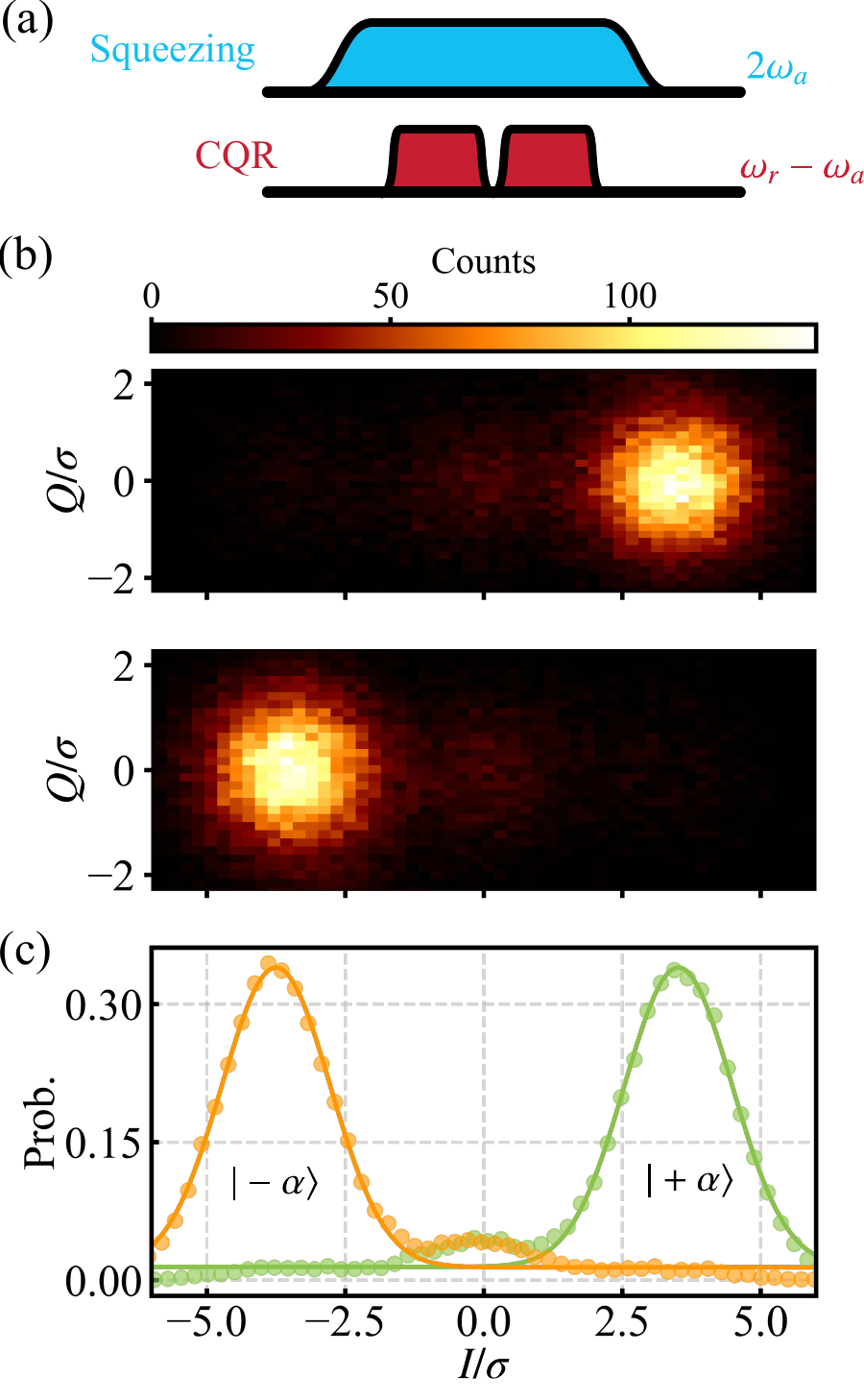}
\caption{\textbf{Calibrating the CQR.} (a). The pulse sequence for measure the QNDness of the CQR. (b). A highly-averaged CQR signal obtained from performing the experiment in (a). (c). The corresponding histogram of the CQR signal in (b), where we perform two consecutive readouts and then postselect the result of the second one based on that of the first one $| \pm \alpha \rangle$. The yellow (green) circles are measured data postselected on measuring $| \pm \alpha \rangle$, where the solid lines are the Gaussian fits.}\label{CQR}
\end{figure}

We perform logical readout on the KCQ following the same pulse sequence that was first experimentally demonstrated in \cite{Grimm2020}. The basic principle of the readout process is that we perform a conditional displacement on the readout cavity conditioned on whether the KCQ is in the state $|+\alpha \rangle$ or $|-\alpha\rangle$. When we drive at $\omega_r - \omega_a$, we obtain the interaction Hamiltonian
\begin{equation}
    H_{\mathrm{CQR}} = g_{\mathrm{CQR}} \left( r^\dagger + r \right) \sigma_z
\end{equation}
where $r$ is the readout mode and $g_{\mathrm{CQR}}$ is linearly proportional to the size of the Kerr-cat $\alpha$ \cite{Grimm2020, Frattini2022}. The intuition of the physical process is that we are driving a beamsplitter between the readout cavity and the stabilized KCQ. This process converts photons in the KCQ to photons in the readout cavity, which exit through the readout pin and then travel up the amplification chain to the ADC of the readout FPGA card. The phase of the signal carries "which-well" information about the KCQ. More detailed treatments of the CQR process can be found in \cite{Grimm2020, Frattini2022}. For all the CQR measurements included in both the main text and the supplement, we drive the beamsplitter for 600 ns and simultaneously integrate the signal coming from the readout cavity for $2.5 \mu$s. Fig. S \ref{CQR} shows the pulse sequence and the result of a readout QNDness measurement, where we perform two consecutive CQRs on the KCQ and post-select the result of the second measurement on the outcomes $|\pm \alpha \rangle$ of the first measurement \cite{Grimm2020}. We measure the QNDness to be $\mathcal{Q} = (p(+\alpha | + \alpha) + p(-\alpha | -\alpha))/2 = 91\%$, where $p(\pm \alpha | \pm \alpha)$ is the probability that the second measurement agrees with the first \cite{Grimm2020, Touzard2019}.

\subsection{Calibrating KCQ gates} \label{Calibrating KCQ gates}

As discussed in the main text, we need two gates on the KCQ to realize universal control: the $\sigma_z (\theta)$ rotation and the $\sigma_x(\pi/2)$ rotation (i.e., the Kerr gate). Both gates were first experimentally demonstrated in \cite{Grimm2020}. In this section, we describe our process for tuning up these gates. \\

\begin{figure}[h!]
\includegraphics[width=1.0\textwidth]{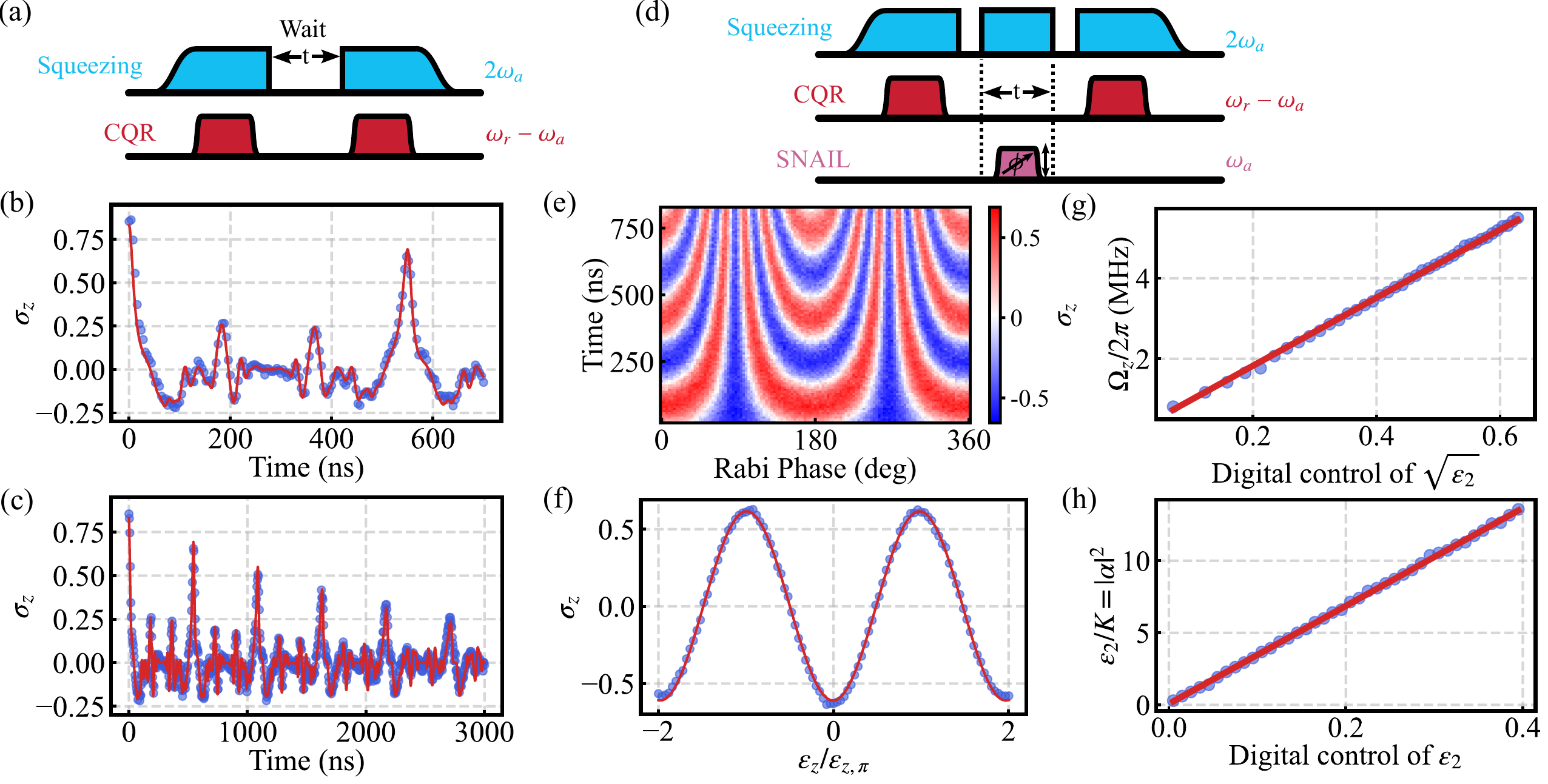}
\caption{\textbf{KCQ gate calibration} (a). Pulse sequence for characterizing the Kerr-refocusing in our SNAIL. We prepare a coherent state $| \pm \alpha \rangle$ in the SNAIL through a CQR measurement, and then switch off the squeezing drive to let the state evolve under Kerr of the SNAIL. We take ``snapshots" of the evolution by switching on the squeezing drive after time $t$ and reading out the state with CQR. (b). The readout signal for the Kerr refocusing measurement. The blue circles are measured data and the red solid line is obtained from a master equation simulation. (c). Longer Kerr refocusing data, obtained by increasing the wait time $t$. The solid line is found using the same master equation simulation as in (b). (d). The control sequence for calibrating a $\sigma_z (\theta)$ rotation on the KCQ. The gaps in the squeezing drive pulse train are the Kerr gates. (e). Rabi oscillations between different cats with different parities $|\mathcal{C}^{\pm}_\alpha \rangle$ as a function of single photon drive phase and duration. (f). Calibrating the amplitude $\epsilon_{z,\pi}$ for a $\sigma_z (\pi)$ rotation, which enables arbitrary $\sigma_z (\theta)$ rotations. (g). KCQ Rabi rate as a function of the square root of the amplitude of our digital control on the squeezing drive. (h).  $\bar{n}_\mathrm{cat}$ in the KCQ as a function of the amplitude of our digital control on the squeezing drive. }\label{Fig.KCQ gates}
\end{figure}

The Kerr-refocusing phenomenon, first demonstrated in \cite{Kirchmair2013}, is the foundation of the discrete $\sigma_x(\pi/2)$ rotation. Under the Kerr-nonlinear Hamiltonian of the SNAIL mode in its rotating frame $H = -Ka^{\dagger}a^{\dagger}aa$, the coherent state $|+\alpha\rangle$ in the SNAIL evolves into the parity-less cat state $|\mathcal{C}^{- i}_\alpha \rangle$ at $t=\pi/K$, and then refocuses into $|-\alpha\rangle$ at $t=2\pi/K$. We can calibrate this refocusing effect using the measurement sequence shown in Fig. S \ref{Fig.KCQ gates}(a). We first prepare a coherent state $|\pm \alpha\rangle $ in the SNAIL via adiabatic squeezing of the SNAIL at $\omega_S=2\omega_a$ and a projective CQR measurement. Then, we abruptly switch off the squeezing drive and let the system evolve under Kerr for different duration $t$, before abruptly switching on the squeezing drive to ``catch" the cat state with the right phase. Lastly, we perform a CQR measurement to determine whether the KCQ has refocused to its initial state. As shown in Fig. S \ref{Fig.KCQ gates}(b), we observe this refocusing peak at $544$ ns, corresponding to a Kerr gate duration of $272$ ns. From this experiment, we extract the Kerr nonlinearity of the SNAIL to be $K/2\pi = 0.93$ MHz. A detailed discussion of the Kerr gate is included in the supplement material of \cite{Frattini2022}.\\

Our Kerr gate quality is limited by $T_1$ errors on the SNAIL. We perform a master equation simulation where the only dissipation on the system is the single photon loss on the SNAIL. We plot the result of the simulation on top of the experimental data in Fig. S \ref{Fig.KCQ gates}(b), and we find excellent agreement between simulation and data, indicating that our Kerr gate fidelity is limited by $T_{1,a}$ of the SNAIL. To further validate our understanding, we increase the duration of the Kerr evolution to include five coherent state refocusing events in Fig. S \ref{Fig.KCQ gates}(c) and we still obtain excellent agreement between our simulation and data.\\

With this Kerr gate, we can now validate the presence of a KCQ in our SNAIL. We replicate the experiment first performed in \cite{Grimm2020}, where we prepare the KCQ in state $|\mathcal{C}^{\pm i}_\alpha \rangle$ and then apply a single-photon drive on the KCQ at $\omega_S/2$. When the single photon drive is in phase with the squeezing drive, it breaks the symmetry of the system and thus lifts the degeneracy between the states $|+ \alpha\rangle $ and $|- \alpha\rangle $, leading to a Rabi rotation of $\sigma_z (\theta)$ around the KCQ Bloch sphere at a rate
\begin{equation}
    \Omega_z = \mathrm{Re}(4\epsilon_z \alpha)
    \label{Eq.KCQ_Rabi}
\end{equation}
to the first order in perturbation theory \cite{Frattini2022}. Note that the "Rabi" terminology comes from the different convention used in \cite{Grimm2020} for the Bloch sphere orientation. When the single-photon drive is out of phase with the squeezing drive, the degeneracy of the potential is restored and the rotation of the system disappears. We demonstrate this phase-dependent rotation rate with a pulse sequence shown in Fig. S \ref{Fig.KCQ gates}(d). We prepare a parity-less cat state $|\mathcal{C}^{\pm i}_\alpha \rangle$ in the SNAIL via a projective CQR measurement and a Kerr gate. Then we drive the SNAIL at $\omega_S/2$ at different phases relative to the squeezing drive for time $t$. Afterward, we read out the KCQ along the $\sigma_y$ quadrature by performing a Kerr gate and a CQR. In the experimental result in Fig. S \ref{Fig.KCQ gates}(e), we see the dependence of the Rabi rate $\Omega_z$ on the phase of the single photon drive $\epsilon_z$, proving that we are generating a bona fide KCQ. \\

Following this demonstration, we use the same pulse sequence to calibrate the $\sigma_z(\pi)$ and $\sigma_z(\pi/2)$ rotation pulses. We use Gaussian pulses ($\sigma$=80 ns, total duration 320 ns) generated by our digital control for this Rabi drive. In the experiment, we calibrate the amplitude of the Gaussian pulses by keeping the single photon drive in phase with the squeezing drive while varying its amplitude $\epsilon_z$, before reading out the KCQ along the $\sigma_y$ axis. An example of such a calibration result is shown in Fig. S \ref{Fig.KCQ gates}(f). We fit the result to a cosine function to calibrate the pulse amplitude $\epsilon_{z,\pi}$ corresponding to a $\sigma_z(\pi)$ rotation. We realize arbitrary $\sigma_z(\theta)$ rotations by varying this amplitude. Combined with the $\sigma_x(\pi/2)$ rotation, this gives us universal control on the KCQ.\\

\subsection{Calibrating the size of the cat $\alpha$} \label{Calibrating the strength of squeezing}

As indicated in Eq. \ref{Eq.KCQ_Rabi}, the Rabi rate of the KCQ is also a function of the size of the cat $\alpha$, which we utilize to calibrate the strength of the squeezing drive \cite{Grimm2020, Frattini2022}. $\alpha$ scales linearly with the digital control amplitude of $\epsilon_2$. If we increase $\epsilon_2$ while holding $\epsilon_z$ constant, we can calibrate the scaling factor by fitting for the slope of the measured KCQ Rabi rates as a function of digital control amplitude. However, this calibration requires an independent measurement of $\epsilon_z$ beforehand, which is done by measuring the Rabi rate of the SNAIL under the drive $\epsilon_z a^{\dagger} + \epsilon_z^*a$ with $\epsilon_2=0$. This gives us $\epsilon_z/2\pi = 0.4$ MHz. With this, we can calibrate $\epsilon_2$ as described. The results are shown in Fig. S \ref{Fig.KCQ gates}(g) where there is a clear linear dependence of the Rabi rate $\Omega_z$ on the square root of $\epsilon_2$ at large drive amplitudes. The slope of this line enables us to calibrate $\bar{n}_{\mathrm{cat}}=|\alpha|^2$ as a function of $\epsilon_2$ as is shown in Fig. S \ref{Fig.KCQ gates}.\\

\subsection{KCQ spectroscopy}

\begin{figure}[h!]
\includegraphics[width=0.8\textwidth]{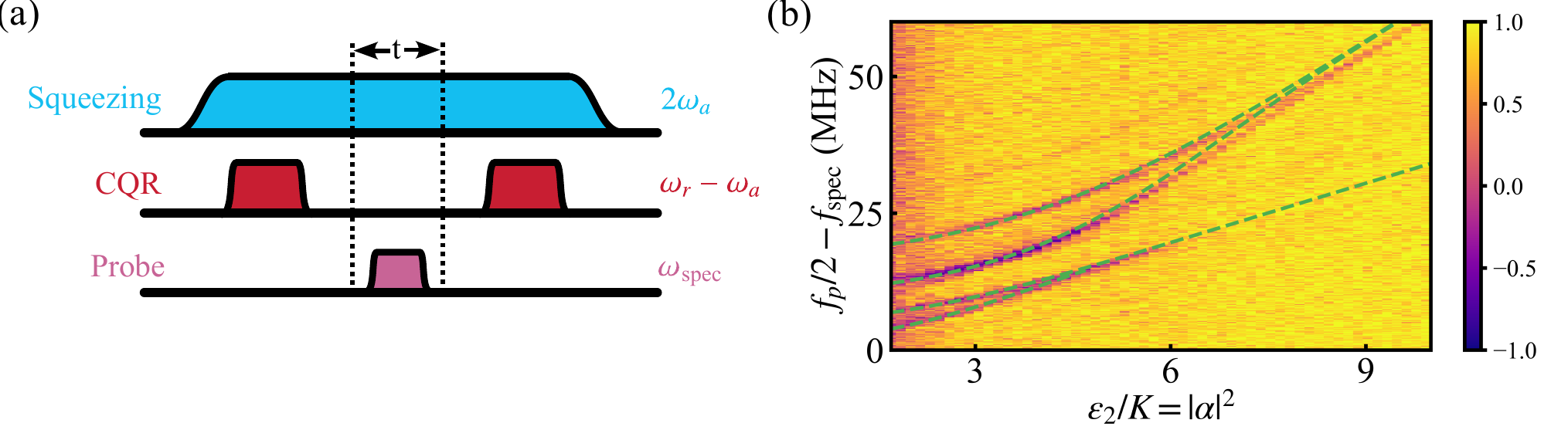}
\caption{\textbf{KCQ spectroscopy} (a). The pulse sequence for KCQ spectroscopy. After preparing a $|\pm \alpha \rangle$ state in the KCQ via CQR, we probe the spectrum of the KCQ with a tone that is detuned from $\omega_a$ for time $t$. When the probe is on resonance with a transition, it will transfer some population to the other well state, which can be resolved using CQR at the end of the measurement sequence. (b). The result of our KCQ spectroscopy experiment. The purple lines are the measured spectroscopy data and the green dashed lines are from diagonalizing the Hamiltonian with parameters independently characterized in other experiments. }\label{Sp_spec}
\end{figure}

In this section, we perform KCQ spectroscopy following the method introduced in \cite{Frattini2022}. This result is useful for performing frequency-selective dissipation (FSD) on the KCQ in later sections. To perform the spectroscopy, we use the pulse sequence shown in Fig. S \ref{Sp_spec}(a). We first prepare $|\pm \alpha \rangle$ in the KCQ by ramping up the squeezing drive and performing a projective CQR measurement. Then, we send in an $800$ ns spectroscopy probe tone with variable frequency $\omega_{\mathrm{spec}}$, near half of the squeezing frequency. If the tone is on resonance with one of the transitions to the excited states, some of the population will be transferred to the other well state $|\mp \alpha \rangle$ via the excited state, which leads to a reduced CQR signal. \\

The spectrum of our KCQ is in Fig. S \ref{Sp_spec}(b). The spectral lines for the first two excited states are performed at a smaller probe tone power than the one used for resolving the second pair of excited states. We stitch the datasets from these separate measurements together as part of the data post-processing. This is the same method used by the authors of \cite{Frattini2022} for their KCQ spectroscopy measurements. It agrees well with our theoretical expectation (in dashed lines). As we increase the squeezing drive amplitude, the spectral lines merge as the states fall inside the wells and become degenerate \cite{Frattini2022}. Fig. S \ref{Sp_spec}(b) tells us that there is more than one level inside the wells when $|\alpha|^2 \gtrsim 5$. \\

\section{Coherence characterization} \label{Coherence characterization}

\subsection{SNAIL coherence characterization} \label{SNAIL coherence characterization}

Owing to the small cross-Kerr $\chi_{ab}/2\pi=2.91$ kHz between the SNAIL and the readout cavity compared with the total linewidth of the readout cavity $\kappa_r/2\pi=0.396$ MHz (Table \ref{Table of parameters}), our dispersive readout of the SNAIL does not have enough SNR for most of our experiments \cite{ReadoutSNR}. Therefore, we are not able to use the standard dispersive readout methods \cite{Blais2004, Wallraff2004} for characterizing the SNAIL coherence. Instead, we adopt a method where we prepare states in the SNAIL, adiabatically ramp up the squeezing drive to map these states to the KCQ Bloch sphere \cite{Grimm2020}, and perform logical readout via CQR. Between the state preparation and the logical readout, the state inside the SNAIL evolves under the dissipation seen by the bare SNAIL, which we use to exact its coherence.\\

\begin{figure}[h!]
\includegraphics[width=0.7\textwidth]{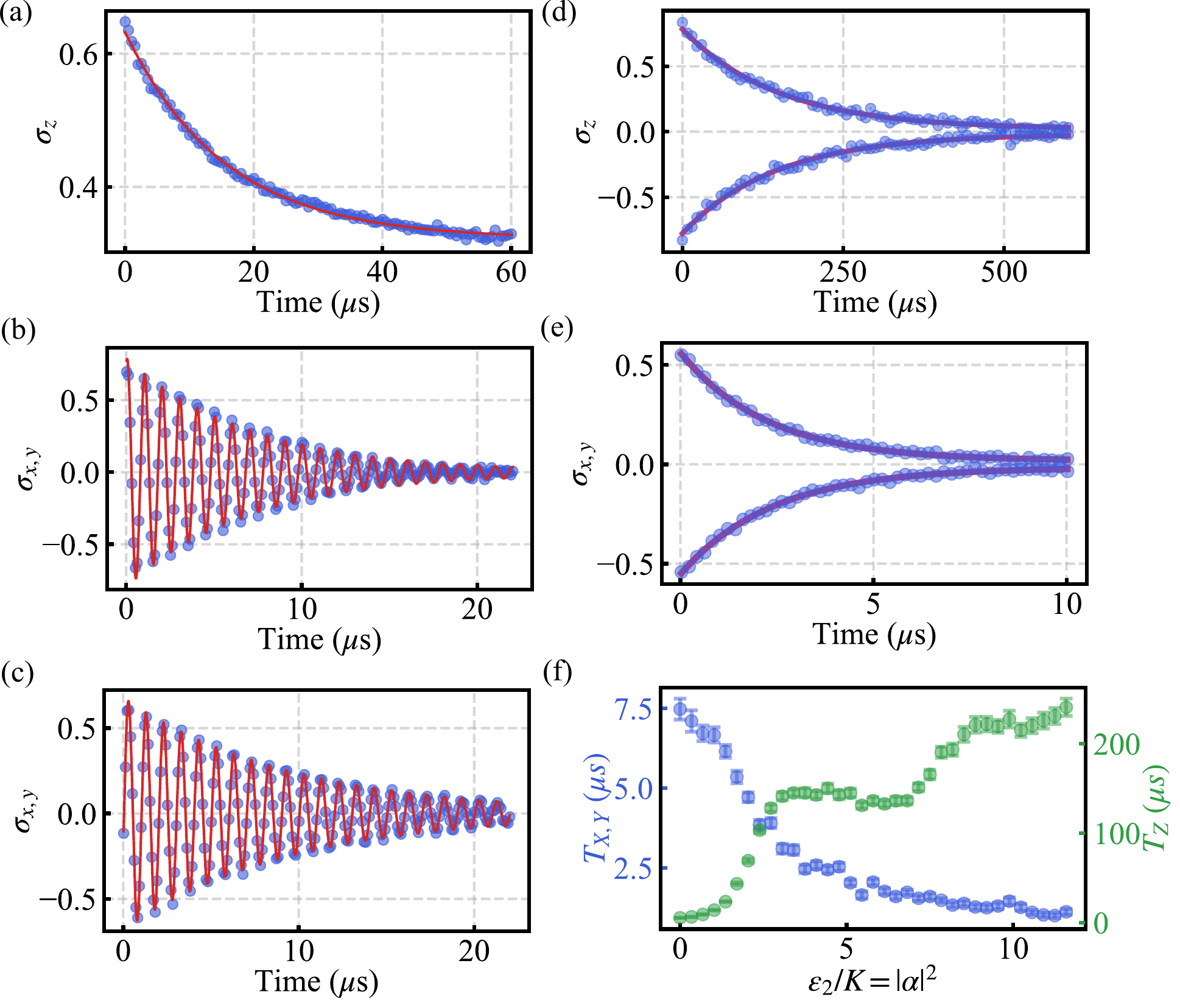}
\caption{\textbf{Coherence characterizations of the SNAIL and the KCQ at $\Phi/\Phi_0 = 0.32$.} (a), (b) and (c): Measurements of the SNAIL $T_{1,a}$, $T_{2,a}$ and $T_{2\mathrm{E},a}$, realized by adiabatically mapping the SNAIL states to the KCQ states and performing CQR. (d) and (e): Measurements of the KCQ lifetimes $T_Z$ and $T_{X,Y}$, corresponding to the coherent states $|\pm \alpha\rangle $ and cat states $|\mathcal{C}^{\pm}_\alpha \rangle$ respectively, using the method in \cite{Frattini2022}. (f). $T_Z$ and $T_{X,Y}$ as functions of $|\alpha|^2$, reproducing the results from \cite{Frattini2022}.}\label{Fig.Sp3}
\end{figure}

To measure the single-photon decay time $T_1$ of the SNAIL, we prepare the excited state $|e\rangle$ in the SNAIL using a square pulse of duration $1.08 \mu$s, such that the frequency distribution of the pulse is a sinc function with a notch at the anharmonicity of the SNAIL. This pulse shaping minimizes leakage to the $|f\rangle$ state of the SNAIL. Then, we idle the SNAIL for time $t$, before adiabatically ramping up the squeezing drive for 2 $\mu$s. We pulse-shape the ramp with a tanh function to minimize leakage. This process maps the YZ plane of the SNAIL qubit Bloch sphere onto the XY plane of the KCQ Bloch sphere so that energy relaxation of the SNAIL from $|e\rangle$ to $|g\rangle$ is mapped to a parity flip of the KCQ (e.g., $|\mathcal{C}^{+}_\alpha \rangle$ to $|\mathcal{C}^{-}_\alpha \rangle$). To read out the parity of the KCQ, we perform a $\sigma_z(\pi/2)$ rotation and a Kerr gate, followed up with CQR. Then we fit the parity as a function of $t$ to extract the energy relaxation rate of the SNAIL $T_{1,a} = 16.0 \pm 0.4 $ $\mu$s (see Fig. S \ref{Fig.Sp3}(a)). \\

To measure the Ramsey decay time of the SNAIL, we prepare a coherent state in the KCQ via a projective CQR measurement. Then we adiabatically map the coherent state $|\pm \alpha\rangle $ of the KCQ to the $|\pm\rangle$ state of the SNAIL. We then idle the SNAIL for time $t$ before mapping it back onto the KCQ for logical readout. To measure the Hahn echo decay time, we perform the same measurement, but we apply a $\pi$ pulse halfway through the idling time to echo out low-frequency noise. We obtain $T_{2R} = 7.2 \pm 0.1 \mu$s and $T_{2E} = 9.5 \pm 0.1 \mu$s, from the measurements shown in Fig. S \ref{Fig.Sp3}(b) and (c), respectively. In these examples, we use a software detuning to artificially rotate our measurement axis at $1$ MHz for better fits.\\

\subsection{KCQ coherence characterization} \label{KCQ coherence characterization}

To characterize the KCQ coherence at our operating point $\Phi/\Phi_0 = 0.32$, we use the same method in \cite{Frattini2022}. We measure the lifetime $T_Z $ of the coherent states $|\pm \alpha\rangle $ by first preparing a coherent state in the KCQ via a projective CQR measurement. Then we idle the KCQ with a constant squeezing drive for duration $t$ and perform CQR. We measure the lifetime of the cat states $T_{X, Y}$ by performing a Kerr gate on the $|\pm \alpha\rangle $ state after the first CQR measurement to prepare $|\mathcal{C}^{\pm i}_\alpha \rangle$ states in the KCQ. After idling for time $t$ with a constant squeezing drive, we perform another Kerr gate to rotate the KCQ so that we are reading it out along the Y axis of the KCQ Bloch sphere. By fitting the population decay shown in Fig. S \ref{Fig.Sp3} (d) and (e), we obtain $T_Z = 147.4 \pm 3.8 \mu$s and $T_{X, Y} = 2.32 \pm 0.04 \mu$s, respectively.\\

We also investigate $T_{\mathrm{X, Y, Z}}$ as a function of the squeezing drive strength to replicate a central result of \cite{Frattini2022}. Essentially, we repeat the measurements described in the previous paragraph on the KCQ at different squeezing drive amplitudes. However, there are two complications with implementing this directly: the Kerr gate fidelity degrades as we increase $|\alpha|^2$ \cite{Frattini2022}, and the CQR needs to be recalibrated for each value of $|\alpha|^2$. To avoid such complications, we perform all the KCQ state preparation and CQR processes at $|\alpha|^2 = 4$. We then change the squeezing drive amplitude for the various idling times required for coherence characterization. Our result in Fig. S \ref{Fig.Sp3}(f) qualitatively agrees with the one in Fig.3(c) of \cite{Frattini2022}. \\

\section{Conditional displacement and cavity displacement calibrations} \label{Conditional displacement and cavity displacement calibrations}

\subsection{CD calibration} \label{CD calibration}

\begin{figure}[h!]
\includegraphics[width=1.0\textwidth]{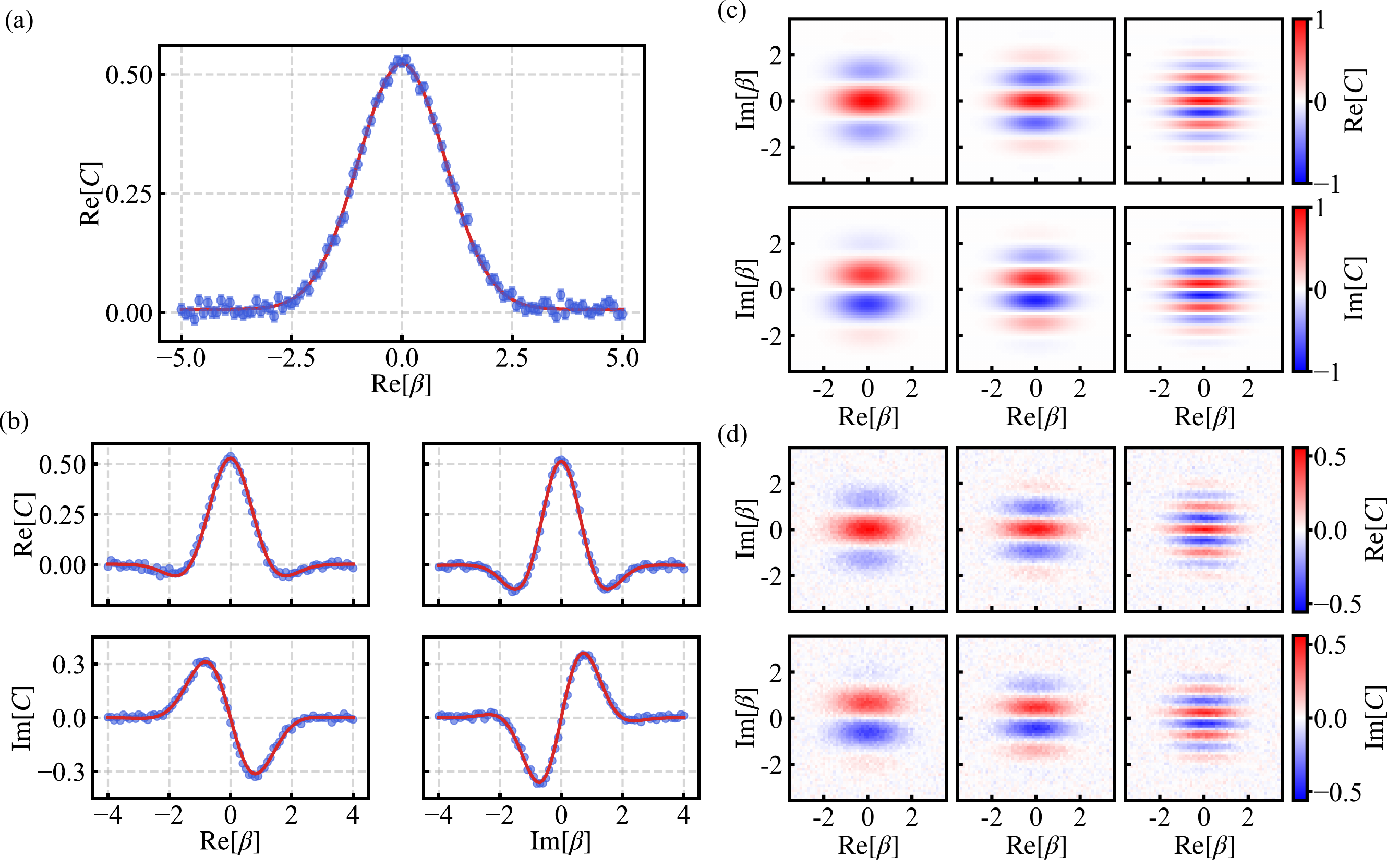}
\caption{\textbf{Conditional displacement and cavity displacement calibrations} (a) CD calibration result, where we take a cut of the vacuum CF tomography along $\mathrm{Re}[\beta]$. The blue circles are measured data and the red solid line is the Gaussian fit. (b) Displacement calibration. After calibrating the CD, we displace the storage cavity and take cuts along $\mathrm{Re}[\beta]$ and $\mathrm{Im}[\beta]$ of the real and imaginary parts of the CF tomography. (c) and (d) Theory predictions and measured data for displacements of 1, 1.5, and 3 (left to right) inside the storage cavity. For each figure, the upper panel is the real part of the CF, while the lower panel is the imaginary part.}\label{Fig.Sp4}
\end{figure}

As a first step towards understanding the coupling between a KCQ and a storage cavity, we need to calibrate the strength of the CD operation. Compared with other cQED implementations where a transmon is dispersively coupled to a high-Q cavity, as in \cite{campagne2020, Touzard2019, Eickbusch2022}, our CD is fundamentally different. As derived earlier, the Hamiltonian of our system, when we are parametrically driving the CD interaction with a time-dependent pulse while stabilizing the KCQ, is given by
\begin{equation}
    H = -Ka^{\dagger 2}a^2 + \epsilon_2 a^{\dagger 2} + \epsilon_2^* a^2 - \chi_{ab} a^{\dagger}a b^{\dagger}b + g_\mathrm{BS}(t)a^\dagger b + g^*_\mathrm{BS}(t) ab^\dagger,
\end{equation}
in the regime where $g_\mathrm{BS}(t) \ll 4\epsilon_2$. We project this Hamiltonian onto the KCQ subspace by using the projector $\mathcal{P}_\mathcal{C} = | \mathcal{C}^+_\alpha \rangle \langle \mathcal{C}^+_\alpha | + | \mathcal{C}^-_\alpha \rangle \langle \mathcal{C}^-_\alpha |$ and obtain
\begin{equation}
    H = \left( g_{\mathrm{CD}}(t)b^\dagger + g_{\mathrm{CD}}^*(t)b \right) \sigma_z -\chi_{ab} \alpha^2 b^\dagger b,
    \label{Eq.CD}
\end{equation}
where $g_{\mathrm{CD}} = \alpha g_{\mathrm{BS}}$ in the large cat limit where $\alpha \gg 1$, as discussed in the main text. The second part of Eq. \ref{Eq.CD} is a Stark shift of the storage cavity due to photons in the KCQ. This Stark shift can be used to characterize $\chi_{ab}$ between the two modes, but for now we focus our attention on the first part of Eq. \ref{Eq.CD}, 
\begin{equation}
    H_{\mathrm{CD}} = \left( g_\mathrm{CD}(t)b^\dagger + g_\mathrm{CD}^*(t)b \right) \sigma_z.
\end{equation}
We see that the CD operation in our setup is parametrically activated by driving a beamsplitter between the SNAIL and the storage, which is fundamentally different than other cQED implementations \cite{Eickbusch2022, campagne2020}. The solution to the time-dependent Schr\"{o}dinger equation $i\hbar \partial_tU =  H_{\mathrm{CD}}U$ is given by
\begin{equation}
    U(t) = \exp\left( -\frac{i}{\hbar} \left( \int_0^t d\tau g_{\mathrm{CD}}(\tau) a^\dagger + \int_0^t d\tau g_{\mathrm{CD}}^*(\tau) a \right) \sigma_z \right),
\end{equation}
which is the CD unitary $U_{\mathrm{CD}} = D(-\beta/2)|-\alpha\rangle \langle -\alpha| + D(+\beta/2)|+\alpha\rangle \langle +\alpha|$ where $\beta = 2\int_0^t d\tau g_{\mathrm{CD}}(\tau)$.\\

We calibrate the strength of our CD operation by measuring the characteristic function (CF) \cite{Fluhmann2020} of the vacuum state $\mathcal{C}(\beta) = \exp(-1/2|\beta|^2)$ \cite{campagne2020}. Experimentally, we measure the CF of the ground state of the storage cavity along $\mathrm{Re}(\beta)$ with the KCQ and fit the results to a Gaussian distribution to extract the calibration scaling factor (Fig. S \ref{Fig.Sp4}(a)). With $|\alpha|^2 = 4$ and a 348 ns pulse (including a 24 ns tanh ramp at both the beginning and end), we obtain $g_\mathrm{CD}/2\pi=6.2 $ MHz. This method relies on the assumption that the storage cavity has a negligible thermal population. Excessive storage thermal population would decrease the width of the Gaussian distribution, leading to a larger scaling factor and thus introducing systematic errors for subsequent calibrations.\\

\subsection{Displacement calibration} \label{Displacement calibration}

Having calibrated the amplitude of our CD drive, we can now calibrate the amplitude of our unconditional displacement drive on the storage cavity. To do so, we prepare a coherent state of unknown amplitude in the storage cavity by displacing it from vacuum, perform CF tomography using the KCQ, and fit the result to its theoretical form given by $\mathcal{C}(\beta)=\exp(-|\beta|^2/2)\exp(\beta\eta^*-\eta\beta^*)$. The best fit for $\eta$ tells us the unknown amplitude of the displacement. In our experiment, rather than performing a full tomography on the displaced state, we sample two cuts along the $\mathrm{Re}[\beta]$ and $\mathrm{Im}[\beta]$ axis, which suffices to determine $\eta$ robustly. \\

After calibrating both the CD and displacement drives, we validate our calibration by preparing coherent states $|\eta=1\rangle, |\eta=1.5\rangle $, and $|\eta=3\rangle$, and performing full CF tomography. We find excellent agreement between theory and experiment, shown in Fig. S \ref{Fig.Sp4}(c) and (d).\\

%%%%%%%%%%%%%%%%%%%%%%%%%%%%%%% starts on 05/09/24 %%%%%%%%%%%%%%%%%%%%%%%%%%%%%%%%%%%%%%%%%%%%%

\subsection{CD rate as a function of cat size} \label{CD rate as a function of cat size}

\begin{figure}[h!]
\includegraphics[width=0.4\textwidth]{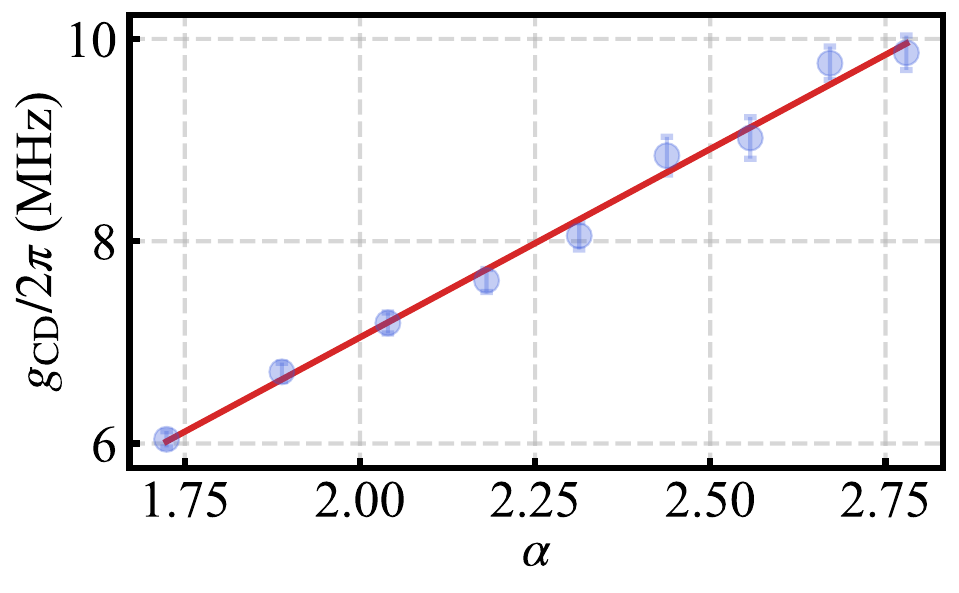}
\caption{\textbf{The rate of CD as a function of cat size $\mathbf{\alpha}$}.}\label{gCD}
\end{figure}

As described earlier, when we drive a beamsplitter between the KCQ and the storage cavity with strength $g_{\mathrm{BS}}$, we obtain the conditional displacement Hamiltonian
\begin{equation}
    H_\mathrm{CD} = \left( g_{\mathrm{CD}}(t)b^\dagger + g_{\mathrm{CD}}^*(t)b \right) \sigma_z,
\end{equation}
where $g_{\mathrm{CD}} = \alpha g_{\mathrm{BS}}$. To validate that $g_{\mathrm{CD}}$ is proportional to cat size $\alpha$, we characterize the CD strength using the method described above at different cat sizes $\alpha$. The measurement result in Fig. S \ref{gCD} displays the predicted linear relationship between $g_{\mathrm{CD}}$ and $\alpha$, with a slope of $g_{\mathrm{BS}} = 1.8$ MHz.\\

There are two important things to note about this measurement. First, we fix $T_{\mathrm{CD}} = 348$ ns so that one unit of CD can be performed with a reasonably small fraction of the digital control amplitude range. This is necessary for our tomography experiments, as we need to use large CDs to perform full CF tomography. If we were to use $T_{\mathrm{CD}} = 100$ ns, we would expect to obtain $g_{\mathrm{CD}}/2\pi \approx 20$ MHz for $|\alpha|^2=4$. Second, there is a trade-off when going to large $\alpha$s to achieve a faster CD. As can be seen from the measurement results in Fig. S \ref{Fig.Sp3}(f), when we increase $\alpha$, $T_{X,Y}$ decreases. This leads to a reduction in the SNR of CF tomography since the coherence of the KCQ becomes comparable to the duration of the measurement ($\sim 1\mu$s). 

\begin{figure}[h!]
\includegraphics[width=0.8\textwidth]{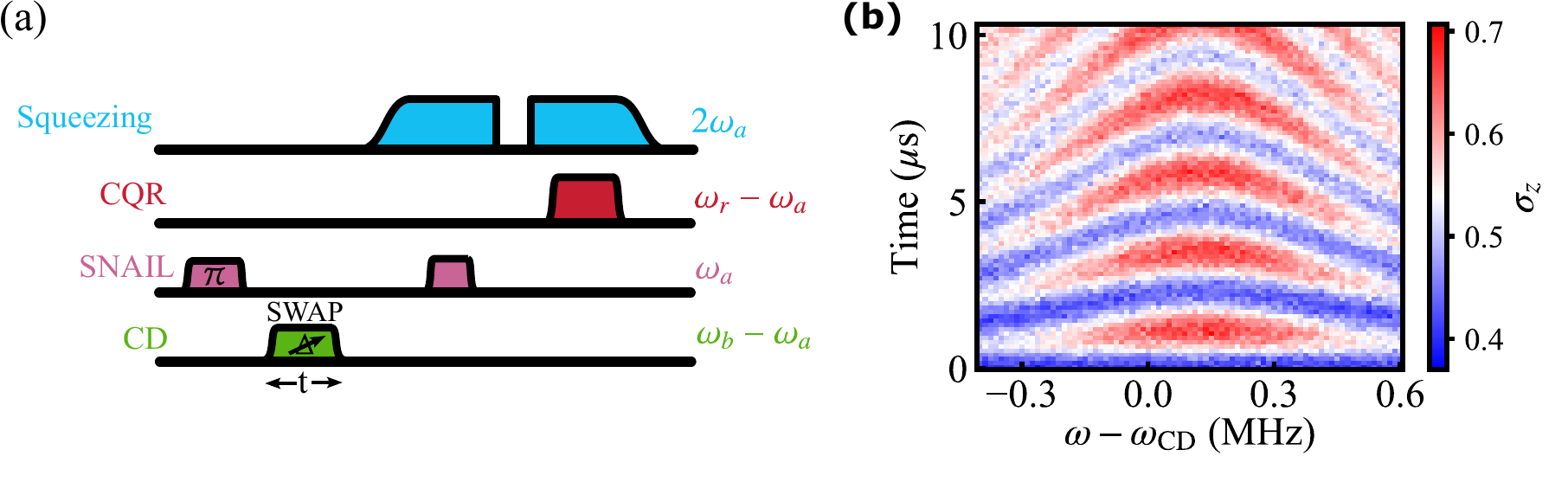}
\caption{\textbf{Calibrating photon swap between a SNAIL and a storage cavity} (a). The control sequence for calibrating the swap. We first prepare the SNAIL in its excited state and then drive a beamsplitter interaction between the SNAIL and the storage at frequency $\omega_b - \omega_a +\Delta$ for time $t$. We then map the SNAIL state to the KCQ and perform a logical readout along the $X$ axis of the KCQ Bloch sphere. (b). The population in the SNAIL as a function of swap frequency and duration. }\label{Fig.Sp_swap}
\end{figure}

\section{Fock state preparation in the storage cavity} \label{Fock state preparation in the storage cavity}

\subsection{Swapping a photon between a SNAIL and a storage cavity} \label{Swapping a photon between a SNAIL and a storage cavity}

\begin{figure}[h!]
\includegraphics[width=0.8\textwidth]{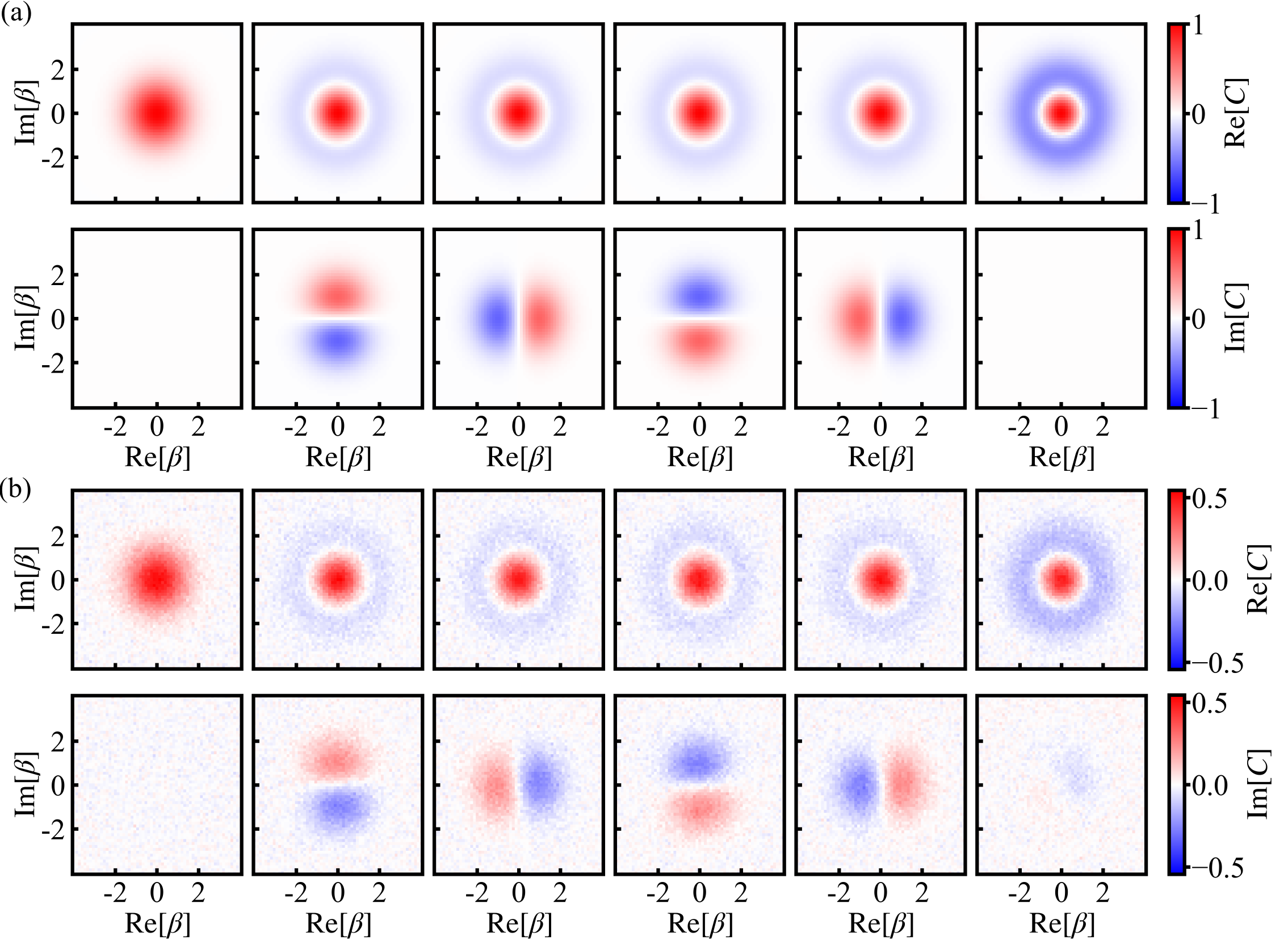}
\caption{\textbf{Storage Fock qubit tomography.} (a) and (b) Theoretical prediction and experimental data of the CF of the cardinal states on the Bloch sphere of the storage Fock qubit. For each figure, the upper panels are the real part of the CF and the lower panels are the imaginary part. From left to right, the states are $|0\rangle$, $|0\rangle + |1\rangle$, $|0\rangle + i|1\rangle$, $|0\rangle - |1\rangle$, $|0\rangle - i|1\rangle$ and $|1\rangle$. Note that the contrast of (b) is much lower than that in (a). This is mostly due to the faulty Kerr gate on the KCQ. }\label{Fig.Sp5}
\end{figure}

As discussed in the main text, to characterize the storage coherence we must prepare cardinal states of the storage Fock qubit spanned by the $|0\rangle$ and $|1\rangle$ states. To do so, we prepare the corresponding state in the SNAIL, and then drive a beamsplitter (BS) interaction between the SNAIL and the storage to swap the single-photon state into the storage cavity. To calibrate this swap, we use the measurement pulse sequence in Fig. S \ref{Fig.Sp_swap}(a). First, we excite the SNAIL from $|g\rangle$ to $|e\rangle$ with the square pulse for $t=1088$ ns. Then we drive the beamsplitter interaction with varying frequency and duration. Next, we adiabatically ramp up the squeezing drive to map the YZ plane of the SNAIL qubit Bloch sphere onto the XY plane of the KCQ Bloch sphere. Finally, we read out the KCQ along the X axis via a $\sigma_z(\pi/2)$ rotation and a Kerr gate. The dataset in Fig. S \ref{Fig.Sp5}(b) displays the standard chevron-like pattern. We fit the cut at the center of the pattern to a damped sine wave and extract the duration $t_{\mathrm{SWAP}} = 1.1 \mu$s and detuning $\Delta = 0.17$ MHz.\\

\subsection{Storage Fock qubit cardinal state tomography} \label{Storage Fock qubit cardinal state tomography}

With a full swap pulse calibrated, we can now prepare all the cardinal states of the storage Fock qubit and validate this preparation via CF tomography. This is the first characterization of a non-Gaussian state with a KCQ. Fig. S \ref{Fig.Sp5}(a) shows the theoretical CFs for all the cardinal states and Fig. S \ref{Fig.Sp5}(b) shows our experimental results. These tomography experiments are performed using the same pulse sequence included in Fig. 2 of the main text. Though the contrast is limited by our Kerr gate fidelity, we find excellent agreement between theory and experiment.\\

\subsection{Storage Fock qubit logical Pauli measurement} \label{Storage Fock qubit logical Pauli measurement}

To measure the coherence of the storage cavity, we measure the decay of the expectation values of the Pauli operators of the storage Fock qubit $\langle X(t) \rangle$, $\langle Y(t) \rangle$ and $\langle Z(t) \rangle$. We use a method adapted from \cite{Koottandavida2023} where we project the displacement operator onto the {$|0\rangle$, $|1\rangle$} subspace in order to express these expectation values in terms of the characteristic function. \\

The single-mode CF for a bosonic mode is defined as 
\begin{equation}
    C(\beta) = \mathrm{Tr}\left( \rho D(\beta) \right)
\end{equation}
where $D(\beta) = \exp(\beta a^\dagger - \beta^* a)$ is the displacement operator for the storage cavity and $\rho$ is the density matrix of the storage cavity. The CF is measured using our CD operation \cite{campagne2020, Eickbusch2022, Fluhmann2020} as described in the main text. The logical codewords of the storage Fock qubit are given by $|+Z\rangle = |0\rangle$ and $|-Z\rangle=|1\rangle$. We define the Fock qubit subspace projector as $P = |0\rangle \langle 0| + |1\rangle \langle 1|$, and we project $D(\beta)$ as
\begin{equation}
    P D(\beta) P = 
    \begin{bmatrix}
    \langle 0| D(\beta) |0\rangle & \langle 0| D(\beta) |1\rangle \\
    \langle 1| D(\beta) |0\rangle & \langle 1| D(\beta) |1\rangle
    \end{bmatrix}
    = e^{-\frac{|\beta|^2}{2}}
    \begin{bmatrix}
    1 & \beta \\
    \beta^* & 1-|\beta|^2
    \end{bmatrix}
    \label{Eq.Akshay_matrix}
\end{equation}
where we use
\begin{equation}
    |\beta\rangle = e^{-\frac{|\beta|^2}{2}} \sum^{\infty}_{n=0} \frac{\beta^n}{\sqrt{n!}} |n\rangle
\end{equation}
and 
\begin{equation}
    \langle 1| D(\beta) |1\rangle = e^{-\frac{|\beta|^2}{2}} \left( 1- |\beta|^2 \right).
\end{equation}
We then equate Eq. \ref{Eq.Akshay_matrix} to the four Pauli matrices, {$X$, $Y$, $Z$ and $I$} and solve for the different $\beta$s. We make the ansatz that $X$ and $Z$ take the form
\begin{equation}
    X = aPD(\beta_1)P + bPD(\beta_2)P
    = ae^{-\frac{|\beta_1|^2}{2}}
    \begin{bmatrix}
    a & a\beta_1 \\
    a\beta_1^* & a-a|\beta_1|^2
    \end{bmatrix}
     +e^{-\frac{|\beta_2|^2}{2}}
    \begin{bmatrix}
    b & b\beta_2 \\
    b\beta_2^* & b-b|\beta_2|^2
    \end{bmatrix} 
    =
    \begin{bmatrix}
    0 & 1\\
    1 & 0
    \end{bmatrix},
\end{equation}
\begin{equation}
    Z = cPD(\beta_1)P + dPD(\beta_2)P
    = e^{-\frac{|\beta_1|^2}{2}}
    \begin{bmatrix}
    c & c\beta_1 \\
    c\beta_1^* & c-c|\beta_1|^2
    \end{bmatrix}
     + e^{-\frac{|\beta_2|^2}{2}}
    \begin{bmatrix}
    d & d\beta_2 \\
    d\beta_2^* & d-d|\beta_2|^2
    \end{bmatrix}
    =
    \begin{bmatrix}
    1 & 0\\
    0 & -1
    \end{bmatrix}.
\end{equation}
To solve this, we make an additional ansatz that $\alpha = -\beta$, $a =-b$, and $c = d$, from which we obtain the solutions 
\begin{equation}
    \begin{split}
        & \beta_1 = \sqrt{2}i, \\
        & \beta_2 = -\sqrt{2}i, \\
        & a = -b = e/2\sqrt{2}i, \\
        & c = d = e/2, \\
    \end{split}
\end{equation}
where $e$ is Euler's number. To measure $I$, we only need to measure the CF at $\beta_0 = 0$, as can be observed from Eq. \ref{Eq.Akshay_matrix}. For $Y$, we need to make a slightly more complicated ansatz involving three displacements
\begin{equation}
\begin{split}
    Y & = fPD(\beta_1)P + hPD(\beta_2)P + jPD(\beta_3)P \\
     &= e^{-\frac{|\beta_1|^2}{2}}
    \begin{bmatrix}
    f & f\beta_1 \\
    f\beta_1^* & f-f|\beta_1|^2
    \end{bmatrix}
     + e^{-\frac{|\beta_2|^2}{2}}
    \begin{bmatrix}
    h & h\beta_2 \\
    h\beta_2^* & h-h|\beta_2|^2
    \end{bmatrix}
    + e^{-\frac{|\beta_3|^2}{2}}
    \begin{bmatrix}
    j & j\beta_3 \\
    j\beta_3^* & j-j|\beta_3|^2
    \end{bmatrix}
    =
    \begin{bmatrix}
    0 & -i\\
    i & 0
    \end{bmatrix}
\end{split}
\end{equation}
to cancel out the real part of the off-diagonal terms while preserving the imaginary part. Here we obtain the solutions
\begin{equation}
    \begin{split}
        & \beta_3 = \sqrt{2}, \\
        & f = h = -ei/2\sqrt{2}, \\
        & j = ei/\sqrt{2}, \\
    \end{split}
\end{equation}
where $e$ is again Euler's constant.\\

All together, we have
\begin{equation}
    \begin{split}
        & I = D(0), \\
        & X = \frac{e}{2\sqrt{2}i} \left[ D(\sqrt{2}i) - D(-\sqrt{2}i) \right], \\
        & Y = \frac{ie}{\sqrt{2}} \left[ D(\sqrt{2}) - D(\sqrt{2}i) - D(-\sqrt{2}i) \right], \\
        & Z = \frac{e}{2} \left[ D(\sqrt{2}i) + D(-\sqrt{2}i) \right]. \\
    \end{split}\label{Eq. fock_sigma}
\end{equation}
These equations tell us that we should measure the CF at these four points, $\beta = {0, \sqrt{2}, \sqrt{2}i, -\sqrt{2}i}$, in order to reconstruct the density matrix of the Fock qubit in the storage cavity. This method can also be used for characterizing cavity coherences for transmon-cavity systems in the weak dispersive regime.\\

\noindent \textbf{Data analysis for Fig. 4(b) in the main text} Monitoring the expectation value of the identity $I$ in Eq. \ref{Eq. fock_sigma} is crucial in our experiments, since it gives us information about changes in the system calibrations and characterization. In the experiment described in Fig. 4 in the main text, we notice that $\langle I (t) \rangle$ increases as a function of time when we are performing the FSD. Through further investigation, we found that the quality of the KCQ $\sigma_x (\pi/2)$ rotation improves as we increase the idling time in the presence of FSD. For good measure, we performed separate measurements to confirm that the storage cavity thermal population stays the same while applying FSD on the KCQ. Critically, this affects the contrast of the CF tomography, and thereby skews the coherence of the storage cavity extracted from this tomography data. To get rid of this effect, we divide the measured Pauli operator expectation values by the measured identity operator expectation values in our data analysis. \\

\section{SNAIL-cavity cross-Kerr characterization} \label{SNAIL-cavity cross-Kerr characterization}

\begin{figure}[h!]
\includegraphics[width=0.8\textwidth]{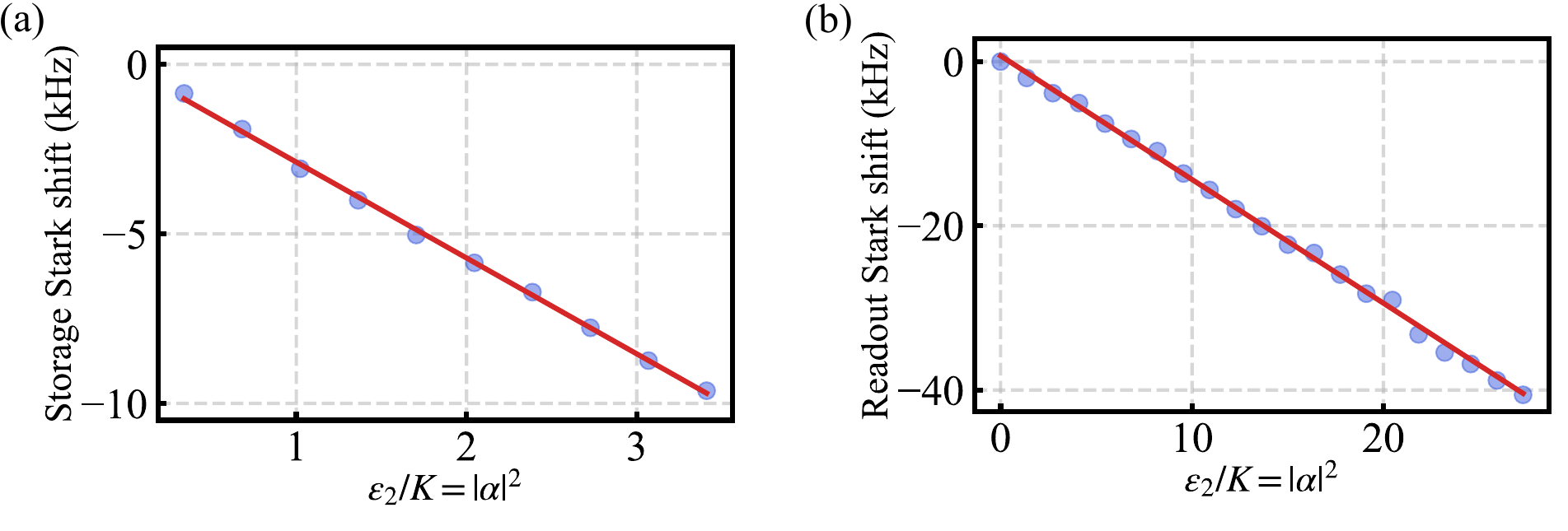}
\caption{\textbf{Cavity Stark shift as a function of $|\alpha|^2$}. (a) and (b) The Stark shifts of the storage and readout cavities as functions of $|\alpha|^2$, respectively. From the linear fits, we can extract $\chi_{ab}/2\pi=2.91 \pm 0.03$ kHz and $\chi_{ar}/2\pi=1.51 \pm 0.01$ kHz.}\label{Stark-shift}
\end{figure}

We calibrate the cross-Kerr between the SNAIL and a cavity mode by measuring the Stark shift of the cavity at a function of the cat size $\alpha$ in the KCQ. The Stark shift is given by the second term in Eq. \ref{Eq.CD}, where the frequency shift of the cavity is linearly proportional to $\alpha$, with the cross-Kerr as the proportionality constant.\\

Specifically, for calibrating $\chi_{ab}$ between the storage cavity and the SNAIL, we measure the rate at which the storage cavity rotates at different values of $\alpha$. To do this, we repeat the experiment in Fig.3 of the main text, where we prepare a single photon state in the storage at a certain $\alpha$ and monitor how fast it is rotating by reconstructing its density matrix as a function of time. We plot the rotation frequency of the storage cavity as a function of $|\alpha|^2$ in Fig. S \ref{Stark-shift}(a), where the slope gives us $\chi_{ab}/2\pi = 2.91 \pm 0.03$ kHz. After obtaining $\chi_{ab}$, we can add in a software detuning at $\chi_{ab}|\alpha|^2$ at different $|\alpha|^2$ for the main experiment in Fig.3 of the main text to cancel out the effects of the deterministic Stark shift for better fit results and better consistency of experimental settings.\\

We also use this method to measure $\chi_{ar}$, the cross-Kerr between the SNAIL and the readout. In this case, we measure the oscillation frequency of the ringdown of the readout cavity at different values of $|\alpha|^2$. For better measurement results, we detune the readout local oscillator by $2$MHz for the measurements. The results in Fig. S \ref{Stark-shift}(b) also display a linear relationship between the readout Stark shift and $|\alpha|^2$ with a slope of $\chi_{ar}/2\pi = 1.51 \pm 0.01$ kHz.\\

\section{SNAIL thermal population measurement}\label{SNAIL thermal population measurement}

\begin{figure}[h!]
\includegraphics[width=0.5\textwidth]{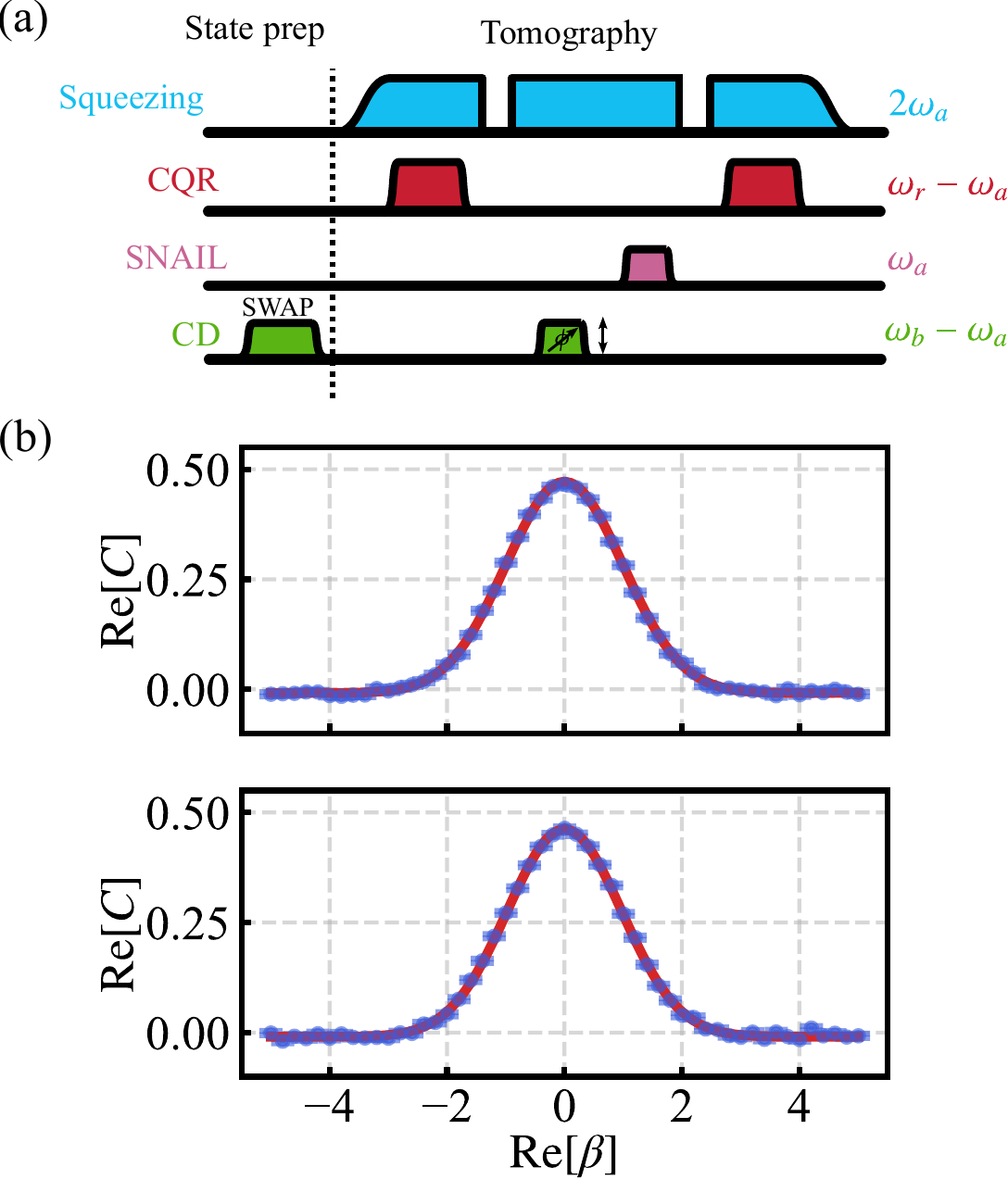}
\caption{\textbf{SNAIL thermal population measurement via a swap test.} (a) The pulse sequence for measuring the thermal population in the SNAIL. We swap the states between the SNAIL and the storage cavity after the system reaches thermal equilibrium and we then perform a CF tomography of the storage along the $\mathrm{Re}[\beta]$ axis. (b). The measurement results of the CF tomography without the swap (upper) and with the swap (lower). The relative change in the width of the Gaussian gives us the thermal population in the SNAIL.}\label{thermal}
\end{figure}

As is discussed in Fig.2 of the main text, we measure the pure dephasing rate of the storage cavity to be $\Gamma_\phi = (5.69 \pm 2.84 \mathrm{ms})^{-1}$ in the presence of the SNAIL (at $\alpha = 0$). To confirm the hypothesis that this dephasing is due to SNAIL heating, we design an experiment to measure the thermal population inside the SNAIL where we take advantage of the existing calibrations and methods. This measurement consists of swapping the states in the SNAIL and the storage after reaching thermal equilibrium, and then performing CF tomography with the KCQ. The width of the Gaussian distribution of the thermal state tells us the thermal population of the SNAIL. The CF of a thermal state is given by 
\begin{equation}
    \mathcal{C}(\beta) = \exp(-(n_\mathrm{th}+\frac{1}{2})|\beta|^2),
\end{equation}
where the width of the Gaussian is given by $\sigma_\mathrm{th}=1/\sqrt{2n_\mathrm{th}+1}$. More thermal population results in a narrower Gaussian distribution.\\

The pulse sequence of this measurement is shown in Fig. S \ref{thermal}. It is almost identical to the experiment shown in Fig.2(a) of the main text, with the distinction that we do not excite the SNAIL before the swap pulse. The results in Fig. S \ref{thermal}(b) show the cut of the storage CF along the $\mathrm{Re}[\beta]$ axis without the swap (upper) and with the swap (lower). From these Gaussian fits, we find that the CF after the swap has a standard deviation that is $0.027 \pm 0.004$ lower than that of the CF without the swap. Assuming negligible thermal population in the storage cavity, we obtain $n_{th} = 0.028 \pm 0.005$ in the SNAIL, giving an effective temperature of $54$ mK. Using the equation from \cite{Wang2019}
\begin{equation}
    \Gamma^{\mathrm{th}}_\phi = \frac{n_\mathrm{th}\kappa_{1a} \chi_{ab}^2}{\kappa_{1a}^2 + \chi_{ab}^2}
\end{equation}
derived in the limit of $n_\mathrm{th} \ll 1$ from \cite{Clerk2007}, where $\kappa_{1a} = 1/T_{1a}$. Plugging in our value for $n_\mathrm{th}$ we obtain $\Gamma_\phi = (7.96 \pm 1.59 \mathrm{ms})^{-1}$, within the errorbar of our independent measurement of the pure dephasing rate in the main text. The errorbar of this result is a lot smaller than the one from the previous method, because this data is much better averaged. This shows that the dephasing of our storage cavity at $|\alpha|^2=0$ is dominated by SNAIL heating.\\

\section{System stability measurement} \label{System stability measurement}

\begin{figure}[h!]
\includegraphics[width=0.8\textwidth]{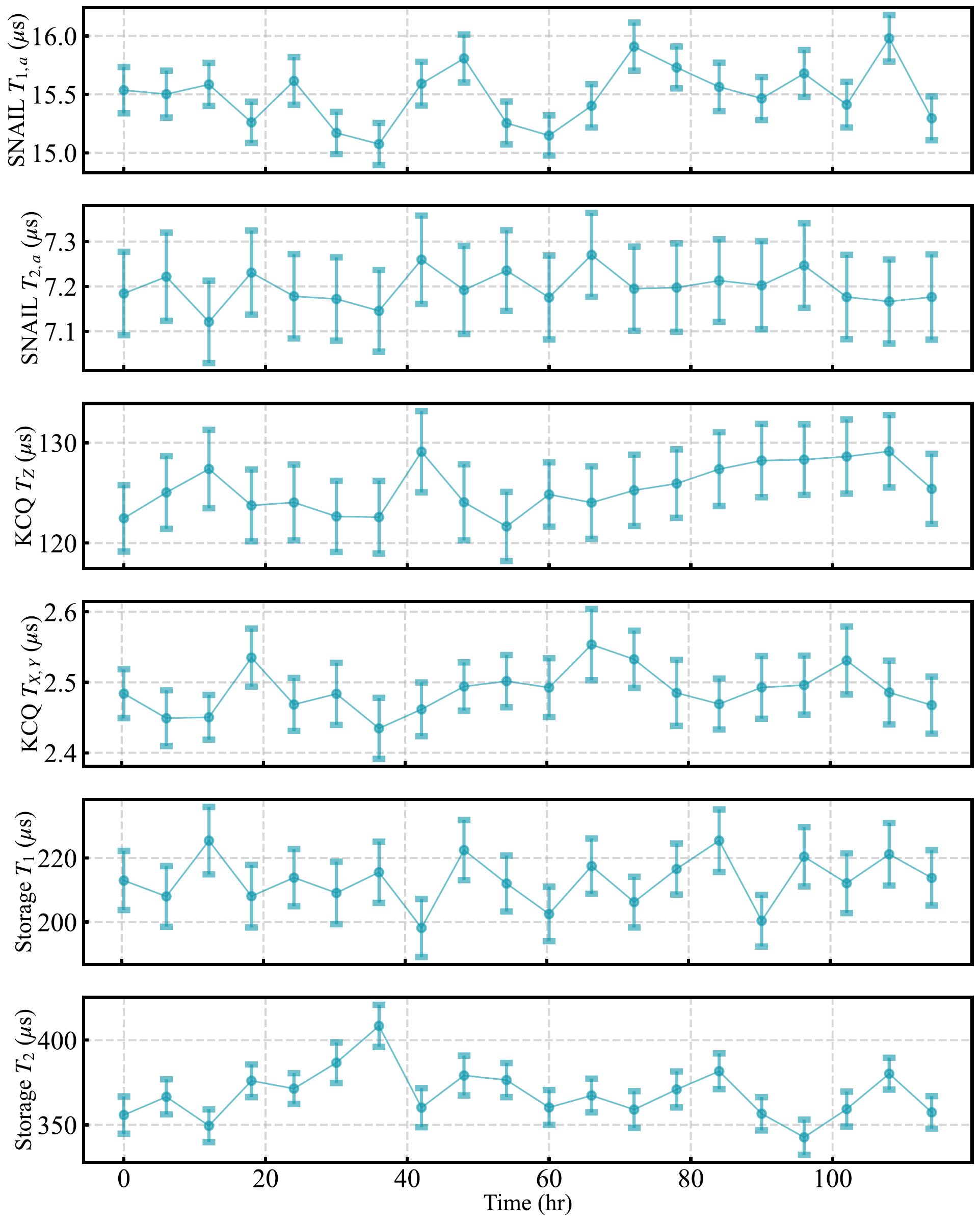}
\caption{\textbf{System stability measurement.} We monitor the coherence of the system over 5 days to test the stability of our coherences. Most of the coherences are fluctuating within $10\%$, while the coherence of the storage cavity is fluctuating within $16\%$. }\label{Sp6}
\end{figure}

The main dataset in Fig.3 of the main text took more than three weeks to complete. We routinely recalibrated the system once a week during this time. This necessitated us to test the long-term stability of our setup. To this end, we ran a one-week-long stability measurement of the system where we monitored the coherences of the bare SNAIL, the KCQ, and the storage cavity over 110 hours. Each set of coherence measurements takes $\sim$2 hours to complete and we did not recalibrate the system in the middle of the test. The measurement result is in Fig. S \ref{Sp6}. Each data point for the storage $T_1$ and $T_2$ is averaged over the same number of shots as the data in Fig.3(b) of the main text. The coherences of the bare SNAIL ($T_{1a}$ and $T_{2a}$) and the KCQ ($T_X$ and $T_Z$) were fluctuating within $10\%$ of their mean values, whereas the storage $T_1$ and $T_2$ were fluctuating within $12\%$ and $16\%$, respectively.\\  

\section{Frequency-selective dissipation on KCQ} \label{Frequency-selective dissipation on KCQ}

\subsection{Basic calibration of the FSD} \label{Basic calibration of the FSD}

\begin{figure}[h!]
\includegraphics[width=0.5\textwidth]{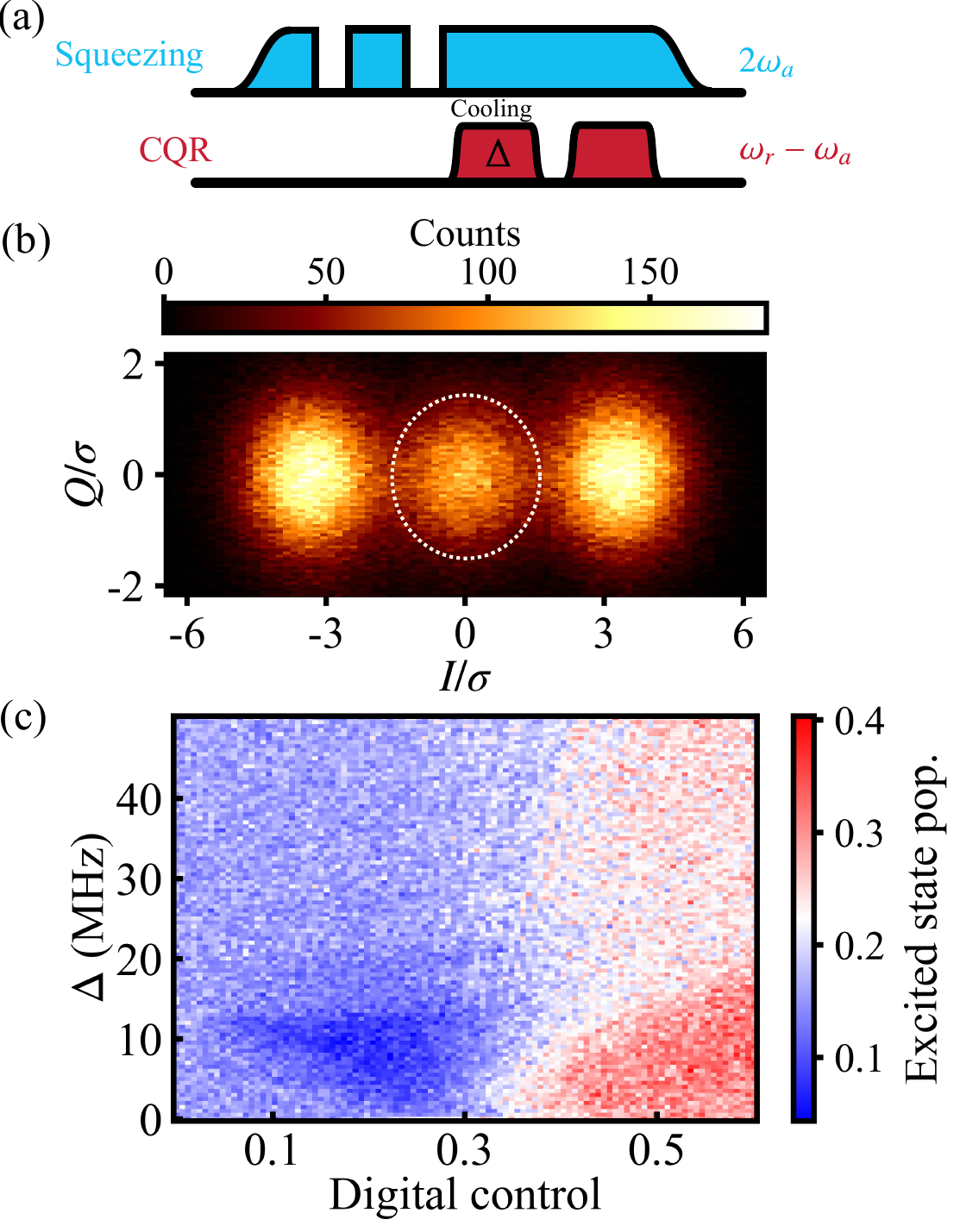}
\caption{\textbf{Calibrating FSD on the KCQ.} (a). The control sequence of the calibration experiment. After preparing a $|\mathcal{C}^{+}_\alpha\rangle$ state in the KCQ with $\alpha=2$, we perform two successive Kerr gates to induce heating in the KCQ. Then we drive the FSD process with different frequencies and amplitudes before we measure the excited state population via CQR. (b). An example of a readout signal of a heated KCQ. The readout blob in the middle, marked by the dashed circle, is the population in the excited states above the wells. (c). KCQ excited state population as a function of cooling amplitude and frequency. The parameters of the point with the lowest population are adopted for our calibrated FSD pulse.}\label{Fig. Cooling}
\end{figure}

In this section, we discuss the calibration of the frequency-selective dissipation (FSD) on the KCQ introduced in the main text. Here, we calibrate the amplitude and frequency of the pulse used to implement the FSD operation. To perform this calibration, we artificially introduce heating to the KCQ, apply the FSD with variable amplitude and detuning, and then perform CQR. We perform this measurement with $|\alpha|^2=4$ for which the first excited state manifold is outside of the well (see Fig. 3 of the main text), such that our CQR has a third blob in the center of the I/Q plane correlated with the excited state population. The goal is to find the FSD parameters that minimize the excited state population.\\

The experimental pulse sequence is shown in Fig. S \ref{Fig. Cooling}(a). At first, we prepare an even parity cat $|\mathcal{C}^{+}_\alpha \rangle$ in the KCQ by adiabatically ramping up the KCQ from the $|g\rangle$ state of the SNAIL \cite{Grimm2020}. Then, we artificially heat the SNAIL by performing two consecutive Kerr gates on the KCQ. As our Kerr gate is mostly limited by single-photon loss on the SNAIL, there is a significant excited state population after the Kerr gates. Subsequently, we perform an FSD by turning on the CQR drive for 5 $\mu$s with variable detuning $\Delta$ and digital control amplitude. We then wait for $400$ ns for the readout cavity to empty itself before our CQR measurement. An example histogram of the measurement is shown in Fig. S \ref{Fig. Cooling}(b), where we indicated the region corresponding to the excited state population. Our measured excited state population as a function of detuning and amplitude is shown in Fig. S \ref{Fig. Cooling}(c). We find it is minimized at $\Delta=12.5$ MHz, which agrees with the gap energy independently measured in KCQ spectroscopy in Fig. S \ref{Sp_spec}, confirming that we are indeed dissipating population from the first pair of excited states of the KCQ.\\

\subsection{FSD on a large Kerr-cat} \label{FSD on a large Kerr-cat}

\begin{figure}[h!]
\includegraphics[width=0.8\textwidth]{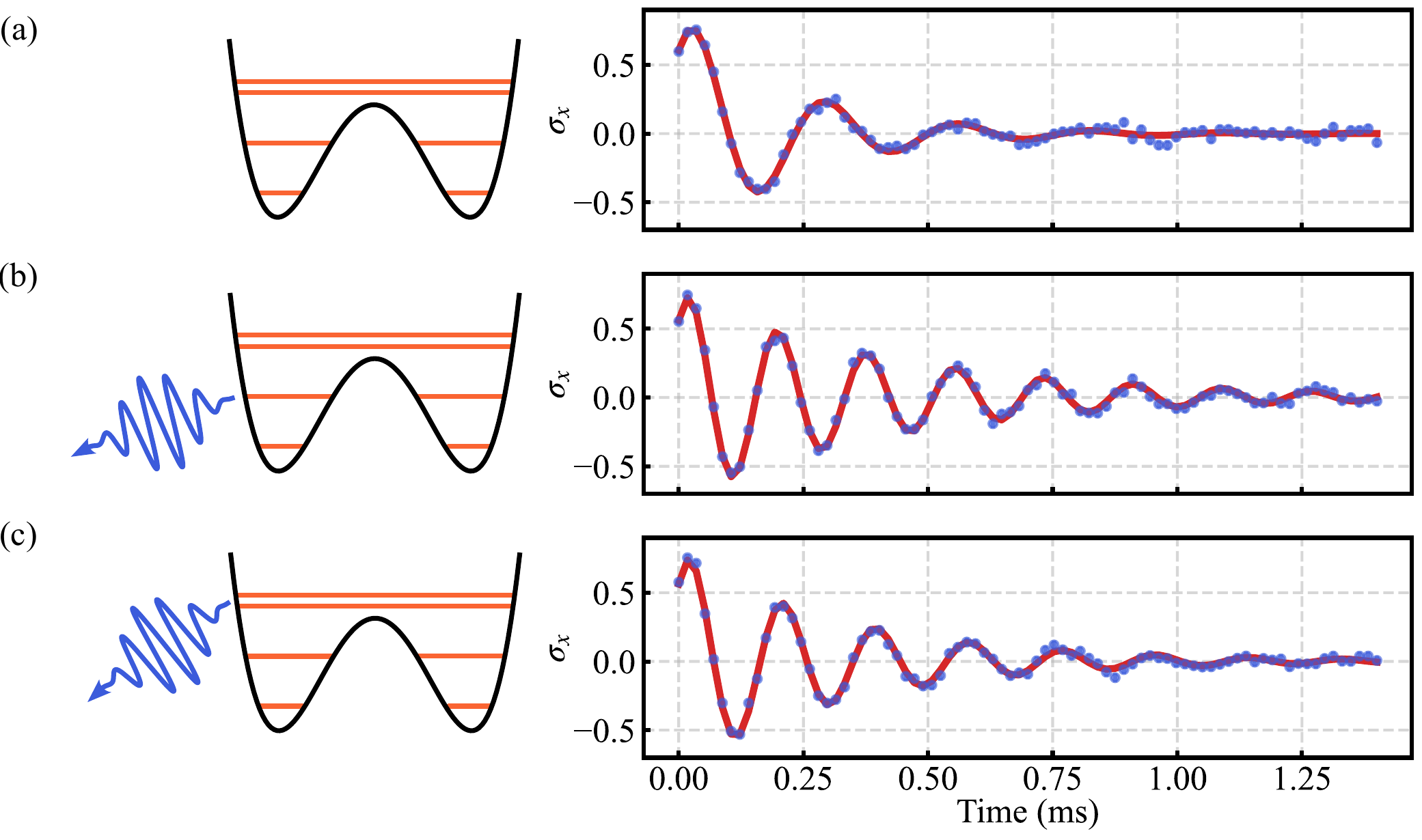}
\caption{\textbf{FSD on a larger cat.} (a). Coherence of the storage in the presence of a $|\alpha|^2=7$ KCQ with $T_2 = 226 \pm 8 \mu$s. (b). Coherence of the storage cavity when dissipating on the first excited state in the KCQ with $T_2 = 433 \pm 12 \mu$s. (c). Coherence of the storage cavity when dissipating on the second excited state in the KCQ with $T_2 = 334 \pm 9$ $\mu$s. The frequencies of the oscillation are different due to the different Stark shifts on the storage cavity from the FSD pulse.}\label{Fig.Sp_largecat}
\end{figure}

We next use the FSD to investigate whether KCQ-induced storage dephasing is due to heating outside of the well or due to heating to the first excited state. To this end, we use a larger Kerr-cat with $|\alpha|^2=7$ where the first pair of excited states is inside the wells and the second pair is outside. We then apply the FSD separately on these levels to see the effects on storage dephasing. To do so, we make use of the KCQ spectroscopy experiment, which tells us that for $|\alpha|^2 = 7$ the first pair of excited states is detuned from the ground state by $20$ MHz while the second pair of excited states is detuned by $38$ MHz. By setting our FSD pulses at these detunings, we are able to target these levels separately. Finally, we use these FSD pulses during the idling time of our storage dephasing measurement (described in Fig.3 of the main text ). \\

The results of these measurements are shown in Fig. S \ref{Fig.Sp_largecat}. In the top panel (a), we apply no FSD on the KCQ, and we obtain $T_2 = 226 \pm 8 $ $ \mu$s. In the middle panel (b), we apply FSD to the first pair of excited states inside the wells, and we obtain $T_2 = 433 \pm 12 $ $\mu$s. This is in excellent agreement with $2T_1 = 432 \pm 17 $ $\mu$s, showing that we have eliminated dephasing to within the precision of our measurement. In the bottom panel (c), we apply FSD on the second pair of excited states just outside of the wells, and we obtain $T_2 = 334 \pm 9$ $\mu$s. This indicates that even heating inside the wells to the first excited state manifold causes storage dephasing, even though these states are degenerate.

%\bibliography{supplement_bib}% common bib file